\documentclass[prb,twocolumn,floatfix,footinbib,bibnotes,longbibliography]{revtex4-1}

\usepackage{graphicx}
\usepackage{epsfig,tabularx,multirow,booktabs}
\usepackage{amsmath,bbm,amssymb,bm,times,mathtools}

\usepackage[colorlinks,linkcolor=blue,citecolor=blue]{hyperref}

\newcommand{\eqn}[1]{
\begin{eqnarray}
	#1
\end{eqnarray}
}

\newcolumntype{C}{>{$}c<{$}}
\AtBeginDocument{
\heavyrulewidth=.08em
\lightrulewidth=.05em
\cmidrulewidth=.03em
\belowrulesep=.65ex
\belowbottomsep=0pt
\aboverulesep=.4ex
\abovetopsep=0pt
\cmidrulesep=\doublerulesep
\cmidrulekern=.5em
\defaultaddspace=.5em
}

\begin{document}
\title{System-environment entanglement phase transitions}

\author{Yuto Ashida}
\email{ashida@phys.s.u-tokyo.ac.jp}
\affiliation{Department of Physics, University of Tokyo, 7-3-1 Hongo, Bunkyo-ku, Tokyo 113-0033, Japan}
\affiliation{Institute for Physics of Intelligence, University of Tokyo, 7-3-1 Hongo, Tokyo 113-0033, Japan}
\author{Shunsuke Furukawa}
\affiliation{Department of Physics, Keio University, Kohoku-ku, Yokohama, Kanagawa 223-8522, Japan}
\author{Masaki Oshikawa}
\affiliation{Institute for Solid State Physics, University of Tokyo, Kashiwa, Chiba 277-8581, Japan}
\affiliation{Kavli Institute for the Physics and Mathematics of the Universe (WPI), University of Tokyo, Kashiwa, Chiba 277-8583, Japan}

\begin{abstract}   
Entanglement in quantum many-body systems can exhibit universal phenomena governed by long-distance properties. We study universality and phase transitions of the entanglement inherent to open many-body systems, namely, the entanglement between a system of interest and its environment. Specifically, we consider the Tomonaga-Luttinger liquid (TLL) under a local measurement and analyze its unconditioned nonunitary evolution, where the measurement outcomes are averaged over. We quantify the system-environment entanglement by the R\'enyi entropy of the post-measurement density matrix, whose size-independent term encodes the universal low-energy physics. We develop a field-theoretical description to relate the universal term to the effective ground-state degeneracy known as the $g$ function in a boundary conformal field theory, and use the renormalization group method to determine its value. 
We show that the universal contribution is determined by the TLL parameter $K$ and can exhibit singularity signifying an entanglement phase transition. 
Surprisingly, in certain cases the size-independent contribution can increase as a function of the measurement strength in contrast to what is na\"ively expected from the $g$-theorem. We argue that this unconventional behavior could be attributed to the dangerously irrelevant term which has been found in studies of the resistively shunted Josephson junction. We also check these results by numerical calculations in the spin-$\frac{1}{2}$ XXZ chain subject to a site-resolved measurement. Possible experimental realization in ultracold gases, which requires no postselections, is discussed. 
\end{abstract}

\maketitle


\section{Introduction}

Understanding universal aspects of entanglement in quantum many-body systems has been a subject of great interest in both condensed matter physics and quantum information science \cite{CP09rev,FE13,ZB19}. A prime example is an entanglement entropy of an interval in a one-dimensional (1D) critical state, which exhibits a universal logarithmic scaling with the coefficient given by the central charge $c$ of the corresponding conformal field theory (CFT) \cite{CH94,VG03,PC04}. Another example is a topologically ordered state, in which the underlying long-range entanglement leads to a universal subleading contribution to the entanglement entropy \cite{HA05,LM06,KA06}. More recently, there has been growing interest in rich and potentially new behaviors of many-body entanglement which are induced by projection measurement \cite{MAR16,NK16,Li19,Zabalo20,GMJ20,GMJ202,ZL20,Jian20,Tang20,Bao20,TX20,FR21,LY21,LA21,SS21,KV21,LTC21,IM21,ZA22,LY23,HO23,HM24,NC22,KJM23,JCH23} or continuous monitoring, i.e., weak nonunitary backaction due to an external environment \cite{WA16,YA18,SJJ18,YA182,XC19,YF20,Goto20,CX20,AO21,GS21,BAl21,TX21,MT22,MTa22,TX22,YK23,GK23,GE23,ZC23}.  All these developments have so far concerned entanglement properties \emph{within} a system of interest, where one partitions a system into a few parts and then considers entanglement between those subsystems.

\begin{figure}[t]
\includegraphics[width=75mm]{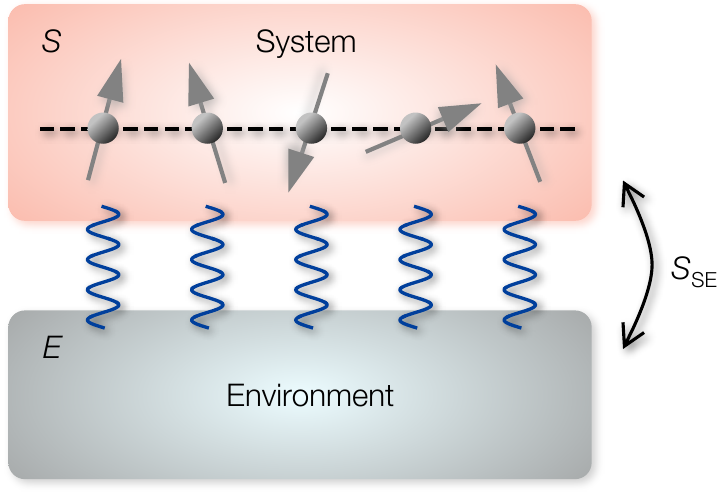} 
\caption{\label{fig_schem}
Schematic figure illustrating the setup. The total Hilbert space consists of an environment $E$ and a one-dimensional many-body system $S$ whose low-energy behavior is described by the Tomonaga-Luttinger liquid. We study universality and phase transitions of the entanglement between $S$ and $E$, which can be diagnosed by the R\'enyi entanglement entropy $S_{SE}$.
}
\end{figure}

The aim of this paper is to reveal yet another universal aspect of entanglement which is inherent to open many-body systems. Specifically, we focus on the entanglement between an entire system and its environment (see Fig.~\ref{fig_schem}), and ask the following questions:
\begin{itemize}
\item[(i)]{Are there phase transitions in the system-environment entanglement, and if so, can they exhibit universal behavior?}
\item[(ii)]{How can one develop a field-theoretical description of the system-environment entanglement?}
\item[(iii)]{Is it possible to analytically calculate the universal contribution to the system-environment entanglement?}
\end{itemize}
Quantum critical states are of particular interest in this context since they are highly entangled states susceptible to external perturbations and expected to exhibit nontrivial long-distance behavior when coupled to the environment. Motivated by this, we address the above questions by considering a class of 1D critical states described as the Tomonaga-Luttinger liquid (TLL) \cite{TSi50,LJM63}. The concept of the TLL provides a unified framework to analyze low-energy physics of various 1D interacting systems ranging from fermionic and bosonic many-body systems to spin chains \cite{TG04,FH81,FH812}. The long-distance correlation, for instance, is characterized by just a single parameter $K$ known as the TLL parameter.

Our main interest lies in the unconditioned nonunitary evolution of the TLL, where the measurement outcomes are averaged over. We answer question (i) in the affirmative way by demonstrating that the TLL subject to a local measurement exhibits a universal entanglement phase transition as a function of the measurement strength.  
Here, the system-environment entanglement is quantified by the R\'enyi entropy of the post-measurement density matrix. 
One of the key findings is that the system-environment entanglement acquires a size-independent universal term $s_0$ that is in general irrational and can exhibit singular changes as the values of $K$ and/or the measurement strength are varied.

We develop a field-theoretical formalism to analyze universality and phase transitions of the system-environment entanglement. Namely, we express the post-measurement density matrix as a vector in a doubled Hilbert space \cite{MDC75,AJ72} and employ the Euclidean path-integral representation \cite{BY23,LJY23}. The resulting field theory is described by the copies of the original theory which corresponds to a $c=1$ CFT in the  case of the TLL. The nonunitary evolution due to the environment is represented as the boundary term acting on the multicomponent (1+1)-dimensional fields. In this description, the universal contribution to the system-environment entanglement can be obtained as the Affleck-Ludwig boundary entropy \cite{AI91}. As such, entanglement phase transitions are described as boundary phase transitions in the corresponding statistical field theory. While the emphasis of our analysis is on the TLL under a local measurement, the present formulation is general and can be used to study the system-environment entanglement in a variety of setups, thereby addressing question (ii).

To analytically obtain the universal contribution as raised in question (iii), we have to approach the problem in two steps. First, we perform renormalization group (RG) analysis to figure out whether or not the boundary action is relevant to long-distance properties. In this way, one can determine which conformal boundary conditions must be imposed on the effective field theory in the infrared (IR) limit. In particular, when we consider local decoherence, the corresponding boundary action is written as local perturbations and can naturally lead to the conformal boundary conditions. 
One of the key challenges in this RG analysis is that one must go beyond the perturbative treatment. This is because the boundary action can have a dangerously irrelevant term, which can be relevant in nonperturbative regions despite being perturbatively irrelevant \cite{MK22,YT23,DR23}. As detailed later, neglect of this term would lead to a result that is at odds with the earlier study \cite{SJM09}.  Second, we construct conformal boundary states consistent with the boundary conditions determined by the RG analysis. To this end, we need a careful treatment of the compactification conditions of the multicomponent fields. Once the correct conditions are identified, the universal constant contribution to the partition function can be obtained by invoking the boundary CFT techniques  \cite{CJ04,JP88,CJ89,HO96,MO97,FS11}. These results are checked by  our numerical calculations in the spin-$\frac{1}{2}$ XXZ chain under a site-resolved measurement.

Before getting into our concrete analyses, let us put the present work in a broader context. First, as discussed later, the universal contribution $s_0$ to the system-environment entanglement can be directly related to the $g$ function in a boundary CFT, which has an interpretation as an effective ground-state degeneracy \cite{AI91,JP00,FD04,CH16,ZY23}.  It is commonly believed that the $g$ function monotonically decreases under RG flows between boundary fixed points, which is often referred to as the $g$-theorem. In other words, when measurement acts as a relevant perturbation, the boundary entropy $s_0$  converges in the thermodynamic limit to a universal value that is less than the initial value of the ultraviolet (UV) theory; as such, one would expect that $s_0$ decreases as the measurement strength is increased. Surprisingly, we find that in certain cases the size-independent contribution $s_0$ can increase as a function of the measurement strength (see Figs.~\ref{fig_rg}(a) and \ref{fig_z_g} below). We speculate that this unconventional behavior originates from nonmonotonic RG flows due to the dangerously irrelevant term that has been discussed in the context of the dissipative quantum phase transition \cite{MK22,YT23,DR23}. 

Second, the present study sharply contrasts with previous studies that have analyzed the TLL influenced by measurement backaction at a single-trajectory level \cite{YA16,YA17nc,DB20,MC21,BM21,YM22,YK22,GSJ23,SX23}. In the latter, nonunitary dynamics is conditioned on the measurement outcomes, and nontrivial effects can appear even in a linear function of the system density matrix such as an expectation value of local observables. This fact has its origin in the nonlocality inherent to quantum measurement \cite{YA18,LY21,MN20}. Meanwhile, the price one must pay is the need of postselecting measurement outcomes, which currently remains a major challenge despite recent efforts \cite{IM21,LTC21,NC22,KJM23,YL23,JCH23}. 
In contrast, our setup requires no postselections while nontrivial effects can be encoded only in a nonlinear function such as the R\'enyi entropy. Notably, recent experimental developments have allowed one to measure such a nonlinear quantity (see, e.g., Ref.~\cite{IR15}); below we will propose a concrete protocol to test our theoretical predictions in ultracold atomic experiments.

Third, the present work also has close connections with the earlier studies in the areas of quantum nanotransport \cite{Kane92,Furusaki93,Bockrath99,Yao99,MO06} and dissipative systems \cite{Leggett87,Caldeira81,Schmidt83,Bulgadaev84,HUR20082208,WU12}. There, a quantum impurity is typically coupled to a bath represented as a collection of bosonic modes. When a bath can be modeled as the Ohmic bath, a canonical transformation can be used to express the oscillator bath in terms of  the TLL \cite{Affleck01}, i.e., a 1D free massless quantum bosonic field. One can use, for instance, precisely the same boundary action as considered in our study to describe the resistively shunted Josephson junction \cite{Schon90}. Naturally, the boundary CFT techniques have found applications to various quantum impurity problems and dissipative systems. The present study demonstrates that these techniques are also useful to study the system-environment entanglement; in particular, our study can provide further insight into the recent discussions about the dissipative quantum phase transition as detailed later.

The remainder of the paper is organized as follows. 
In Sec.~\ref{sec:gen}, we present a general formulation to describe the system-environment entanglement within the field-theoretical framework.
In Sec.~\ref{sec:tll}, we introduce a model of the TLL under a local measurement. 
In Sec.~\ref{sec:rg}, we perform both nonperturbative and perturbative RG analyses of the boundary action and identify the conformal boundary conditions in the IR limit that will be necessary in the boundary CFT analysis. 
In Sec.~\ref{sec:bcft}, we employ the boundary CFT techniques to determine the value of the universal contribution to the system-environment entanglement.
In Sec.~\ref{sec:num}, we present the numerical analysis of the spin-$\frac{1}{2}$ XXZ chain under a site-resolved measurement and demonstrate a consistency with the analytical results obtained in the preceding sections.  
In Sec.~\ref{sec:exp}, we briefly discuss a possible way to test our theoretical predictions in ultracold atomic experiments.
In Sec.~\ref{sec:sum}, we give a summary of our results and suggest several directions for future investigations.

\section{General Formulation\label{sec:gen}}
\subsection{System-environment entanglement\label{subsec:see}}
We consider the Hilbert space that consists of a system $S$ and its environment $E$. Suppose that the initial state is prepared in the product state $\hat{\rho}_{S}\otimes\hat{\rho}_{E}$. The unitary operator $\hat{U}$  is acted on the total Hilbert space to generate the entanglement between $S$ and $E$:
\eqn{
\hat{\rho}_{SE}=\hat{U}\left(\hat{\rho}_{S}\otimes\hat{\rho}_{E}\right)\hat{U}^{\dagger}.
}
The system-environment entanglement can then be evaluated by the R\'enyi entanglement entropy,
\eqn{S_{SE}^{(n)}=\frac{1}{1-n}\log{\rm tr}\left[\hat{\rho}_{{\cal E}}^{n}\right],}
where we introduce the reduced system density matrix by
\eqn{\hat{\rho}_{{\cal E}}={\cal E}\left(\hat{\rho}_{S}\right)\equiv{\rm tr}_{E}\left[\hat{\rho}_{SE}\right].}
Here, we take the partial trace over $E$, and $\cal E$ denotes the corresponding completely positive and trace preserving (CPTP) map that describes an effective evolution of the system, which is in general nonunitary. From now on we focus on the case of $n=2$, which corresponds to the purity of the system, and abbreviate the label $n$ for the sake of notational simplicity: 
\eqn{\label{secondrenyi}
S_{SE}=-\log{\rm tr}\left[\hat{\rho}_{{\cal E}}^{2}\right].
}

We consider the situation in which the initial state of the system is given by $\hat{\rho}_S=|\Psi_0\rangle\langle\Psi_0|$ with $|\Psi_0\rangle$ being a 1D critical ground state. When $S$ and $E$ locally interact with each other and exhibit only short-range correlations, the leading contribution to $S_{SE}$ is simply given by the term that scales with the size of the system $L$ (cf. Fig.~\ref{fig_schem}). Consequently, we expect the relation
\eqn{\label{sse}
S_{SE}=s_{1}L-s_{0}+{\rm o}(1).
}
As discussed later, the coefficient $s_1$ is nonuniversal since it depends on microscopic details and is sensitive to a choice of the UV cutoff $\Lambda_0$ in the effective field theory. Namely, the leading contribution originates from high-energy fluctuations and does not reflect low-energy universal properties.   In fact, it is the size-independent term $s_0$ that characterizes universal long-distance properties of the system-environment entanglement. 

The universal contribution $s_0$ allows us to diagnose whether or not the nonunitary mapping $\cal E$ is a relevant perturbation to the long-distance behavior of $\hat{\rho}_S$. When $s_0$ vanishes, $\cal E$ is irrelevant in the RG sense and the low-energy degrees of freedom are effectively decoupled from the environment. In contrast, nonzero $s_0$ indicates that $\cal E$ is a relevant perturbation; in this case, the system-environment coupling typically flows to the strong-coupling limit. Consequently, the system gets strongly entangled with the environment in the IR limit. Interestingly, when the initial critical state $\hat{\rho}_S$ is the TLL as discussed below, we find that $s_0$ can continuously vary depending on the TLL parameter $K$ and  exhibit singularity signifying an entanglement phase transition as a function of the system-environment coupling strength. 

We note that these nontrivial phenomena can be detected only by a quantity that is nonlinear in $\hat{\rho}_{\cal E}$. To illustrate this, it is useful to express the CPTP map by the product of the local maps and employ the Kraus representation \cite{KK71},
\eqn{\label{krausproduct}
{\cal E}=\prod_j{\cal E}_j,\;\;\;{\cal E}_j(\cdot)=\sum_{m}\hat{K}_{m,j}(\cdot)\hat{K}_{m,j}^{\dagger}.
}
Here, the Kraus operators $\hat{K}_{m,j}$ act on site $j$ and satisfy  $\sum_{m}\hat{K}_{m,j}^{\dagger}\hat{K}_{m,j}=\hat{I}$ with $\hat{I}$ being the identity operator. 
Using the dual mapping ${\cal E}^{*}=\prod_{j}{\cal E}_{j}^{*}$ with ${\cal E}^{*}_j(\cdot)\equiv\sum_{m}\hat{K}_{m,j}^{\dagger}(\cdot)\hat{K}_{m,j}$, an expectation value of a local observable $\hat{O}$ with respect to $\hat{\rho}_{\cal E}$ can be expressed by ${\rm tr}[\hat{O}\hat{\rho}_{{\cal E}}]={\rm tr}[{\cal E}^{*}(\hat{O})\hat{\rho}_{S}]$. The latter is nothing but an expectation value of another local observable ${\cal E}^*(\hat{O})$ with respect to $\hat{\rho}_S$, which is not expected to exhibit singular behavior as the Kraus operators are continuously varied. Thus, a linear function of $\hat{\rho}_{\cal E}$ cannot be used to detect the entanglement phase transitions described above.

\subsection{Effective field theory in a doubled Hilbert space\label{subsec:path}}
To develop a field-theoretical approach to analyzing the system-environment entanglement, we first employ the Choi-Jamiolkowski isomorphism \cite{MDC75,AJ72}. Specifically, we rewrite the reduced system density matrix $\hat{\rho}_{\cal E}$ as a vector $|\rho_{\cal E}\rparen$ in a doubled Hilbert space,  
\eqn{\label{rhoed0}
|\rho_{{\cal E}}\rparen=\prod_j\Bigl(\sum_{m}\hat{K}_{m,j}\otimes\hat{K}_{m,j}^*\Bigr)|\rho_{S}\rparen,
}
where a CPTP map $\cal E$ is expressed as an operator acting on the doubled initial pure state $|\rho_{S}\rparen=|\Psi_0\rangle\otimes|\Psi_0^*\rangle$. The positivity of $\cal E$ allows one to 
 write $|\rho_{{\cal E}}\rparen$ in the exponential form,
\eqn{\label{rhoed}
|\rho_{{\cal E}}\rparen=\exp\Bigl(-\mu\sum_{ja}\hat{k}_{ja}\otimes\hat{\tilde{k}}_{ja}\Bigr)|\rho_{S}\rparen,
}
where $\hat{k}_{ja}$ and $\hat{\tilde{k}}_{ja}$ are certain local operators and $\mu>0$  is a dimensionless coefficient that characterizes the system-environment coupling strength or the measurement strength. 

We next employ the path-integral representation and formulate the problem in terms of an effective field theory. To this end, we describe the matrix elements of the doubled density matrix  $|\rho_{{\cal E}}\rparen\lparen\rho_{{\cal E}}|$ by using the Euclidean path integral of the two (1+1)-dimensional scalar fields $\phi$ and $\tilde{\phi}$ \footnote{Here, the fields live on a surface of size $L\times\beta$, where $L$ is the spatial length and the inverse temperature $\beta$ will eventually be taken to infinity to reach the zero-temperature limit. In discussing the transition amplitude in Eq.~\eqref{pp1}, the fields are fixed on the two edges at $\tau=0$ and $\beta$. We express the locations of these edges by $\tau=0^+$ and $0^-$, respectively, as we later glue these edges together to calculate the trace in Eq.~\eqref{ssez}; see also Fig.~\ref{fig_cft2}.},
\eqn{
\lparen\phi'(x),\tilde{\phi}'(x)|\rho_{{\cal E}}
\rparen\!\!\!\!&&\!\!\!\lparen
\rho_{{\cal E}}|\phi''(x),\tilde{\phi}''(x)\rparen \nonumber \\ \nonumber
&=&\!\frac{1}{Z_{{\cal I}}}\int_{(\phi,\tilde{\phi})_{\tau=0^{+}}=(\phi'',\tilde{\phi}'')}^{(\phi,\tilde{\phi})_{\tau=0^{-}}=(\phi',\tilde{\phi}')}{\cal D}\phi{\cal D}\tilde{\phi}\,e^{-{\cal S}_{{\rm tot}}^{{\cal E}}[\phi,\tilde{\phi}]}.\\
\label{pp1}} 
The total action ${\cal S}^{\cal E}_{\rm tot}$ is given by
\eqn{\label{stotact}
{\cal S}_{{\rm tot}}^{{\cal E}}[\phi,\tilde{\phi}]\equiv{\cal S}_{0}[\phi]+{\cal S}_{0}[\tilde{\phi}]+{\cal S}_{{\cal E}}[\phi,\tilde{\phi}],
}
where ${\cal S}_0$ is the bulk action of the ground state $|\Psi_0\rangle$  defined as an integral over the spatial coordinate $x$ and the imaginary time $\tau$, and $Z_{{\cal I}}$ is the partition function of the two decoupled copies of scalar fields, i.e., $Z_{{\cal I}}=(Z_{0})^{2}$ with $Z_{0}=\int{\cal D}\phi\,e^{-{\cal S}_{0}[\phi]}$. For instance, in the TLL considered later, ${\cal S}_0$ will be given by a $c=1$ CFT. Meanwhile, ${\cal S}_{{\cal E}}$ represents the effect of the state changes due to the environment. For the sake of simplicity, we assume that the Kraus operators are diagonal in terms of the field variables. Thus, from Eq.~\eqref{rhoed}, ${\cal S}_{{\cal E}}$ can be written as a boundary term acting on the $\tau=0$ line \footnote{Precisely speaking, when Eq.~\eqref{seact} is to be used in the transition amplitude in Eq.~\eqref{pp1}, it must be understood that the term $k_{ja} \tilde{k}_{ja}$ acts on the $\tau=0^-$ edge while its complex conjugate acts on the $\tau=0^+$ edge. These edges are glued together and turn into the single $\tau=0$ line when discussing the trace in Eq.~\eqref{ssez}.},
\eqn{\label{seact}
{\cal S}_{{\cal E}}[\phi,\tilde{\phi}]\!=\!\mu\!\int dxd\tau\delta(\tau)\sum_{ja}\!\left(k_{ja}[\phi]\tilde{k}_{ja}[\tilde{\phi}]+{\rm c.c.}\right),
}
which induces the interaction between the two copies of scalar fields at the boundary. 

We note that the boundary action ${\cal S}_{\cal E}$  should satisfy several conditions and cannot be chosen arbitrarily. First, ${\cal S}_{{\cal E}}$ should be nonnegative because of the positivity of $\cal E$. Second, the normalization condition of the Kraus operators leads to $\int d\phi\sum_{ja}k_{ja}[\phi]\tilde{k}_{ja}[\phi]=0$, which is necessary to ensure the normalization condition in the weak-measurement limit $\mu\to 0$. Third, if the diagonal elements of $\sum_{ja}\hat{k}_{ja}\otimes\hat{\tilde{k}}_{ja}$ vanish, the boundary action is subject to an additional constraint $ {\cal S}_{{\cal E}}[\phi,\phi]=0$, which will be the case in our examples below.

The nonunitary evolution represented by the temporal defect ${\cal S}_{\cal E}$  can induce a boundary phase transition, which manifests itself as a singular change of the universal contribution $s_0$ in the system-environment entanglement $S_{SE}$ in Eq.~\eqref{sse}.  
In the path-integral representation, $S_{SE}$ can be obtained by the ratio between the two partition functions 
\eqn{\label{ssez}
S_{SE}=-\log{\rm tr}\left[|\rho_{{\cal E}}\rparen\lparen\rho_{{\cal E}}|\right]=-\log\frac{Z_{{\cal E}}}{Z_{{\cal I}}}.
}
We note that the fields in $Z_{\cal I}$ obey the following constraint: 
\eqn{(\phi,\tilde{\phi})_{\tau=0^-}=(\phi,\tilde{\phi})_{\tau=0^+}.\label{pericon}} 
Meanwhile, ${Z_{{\cal E}}=\int{\cal D}\phi{\cal D}\tilde{\phi}\,e^{-{\cal S}_{{\rm tot}}^{{\cal E}}[\phi,\tilde{\phi}]}}$ is the partition function of the two copies subject to Eq.~\eqref{pericon} and possible additional constraints due to the boundary action ${\cal S}_{\cal E}$. 
As shown below, when these constraints lead to certain conformally invariant boundary conditions, the boundary CFT techniques allow us to explicitly calculate each of the partition functions as  \footnote{In fact, both the linear and the constant term vanish for $\xi={\cal I}$ in Eq.~\eqref{zxi} since the partition function $Z_{\cal I}$ is defined for two decoupled tori, which have no boundary; see Fig.~\ref{fig_cft2}. We explicitly show these vanishing terms to emphasize the relative change due to the boundary action. Specifically, as seen in Eq.~\eqref{univcont}, the constant contribution to the system-environment entanglement is related to the relative change in the boundary entropy. We also note that, in general, a term proportional to the spacetime area $\beta L$, i.e., a bulk contribution, adds to Eq.~\eqref{zxi}. This bulk term is insensitive to the boundary condition and cancels between $\xi={\cal E}$ and $\xi={\cal I}$ in Eq.~\eqref{ssez}.}
\eqn{\label{zxi}
\log Z_{\xi}=b_{\xi}L+\log g_\xi+{\rm o}(1),\;\;\xi\in\{{\cal I},{\cal E}\}
}
where $b_\xi$ is a cutoff-dependent nonuniversal coefficient, and $g_\xi$ is a UV-independent universal contribution known as the $g$ function \cite{AI91}.  The latter can be interpreted as the effective ground-state degeneracy, which in general takes a noninteger value determined from the conformal boundary states (cf. Eq.~\eqref{gdeg} and the related discussions in Sec.~\ref{sec:bcft}). 
Comparing Eq.~\eqref{sse} with Eqs.~\eqref{ssez} and \eqref{zxi}, the size-independent universal contribution $s_0$ can be directly related to the $g$ functions via
\eqn{\label{univcont}
e^{s_{0}}=\frac{g_{{\cal E}}}{g_{{\cal I}}}.
}
A nonzero value of $s_0$ then indicates that ${\cal S}_{\cal E}$ is a relevant perturbation, which imposes nontrivial conformal boundary conditions and alters the value of the $g$ function. Physically, this means that the system-environment interaction is relevant in the sense that its influence on the entanglement survives even in the low-energy limit. 
We have thus mapped the problem of characterizing the universality of the system-environment entanglement to the problem of identifying conformal boundary conditions in the IR limit. 

We note that the $g$ function is known to play a similar role in boundary RG flows as the central charge $c$ does in bulk RG flows. Namely, the $g$-theorem states that the $g$ function should monotonically decrease under RG flows, leading to $g_{\cal E}<g_{\cal I}$ provided that the boundary perturbation is relevant \cite{AI91,FD04,CH16}. As such, one might be tempted to conclude that $s_0$ must be less than or equal to zero. We find, however, that this is not always the case. Below we will present a simple example where $s_0$ can take a strictly positive value due to the dangerously irrelevant term, while we will argue that this behavior can be still consistent with the $g$-theorem. 

\section{Tomonaga-Luttinger liquid influenced by a local measurement \label{sec:tll}}
As a concrete example, we study the case when a critical state $|\Psi_0\rangle$ is described by the TLL realized as the ground state of the Hamiltonian,
\eqn{
\frac{\hat{H}}{\hbar}=\int dx\frac{v}{2\pi}\left[\frac{1}{K}\left(\partial_{x}\hat{\phi}\right)^{2}+K\left(\partial_{x}\hat{\theta}\right)^{2}\right],
} 
where $v$ is the velocity, $K$ is the TLL parameter, and $\hat{\phi}(x)$ and $\hat{\theta}(x)$ are the bosonic field operators satisfying the commutation relation $[\hat{\phi}(x),\partial_{x'}\hat{\theta}(x')]=i\pi\delta(x-x')$.  A smaller $K$ means stronger correlations in density fluctuations $\partial_x\hat{\phi}$ and weaker correlations in phase fluctuations $\partial_x\hat{\theta}$. 
We choose the unit $\hbar=v=1$ below and impose the periodic boundary conditions throughout this paper.

When the TLL is realized in a gapless spin-$\frac{1}{2}$ antiferromagnetic XXZ chain, each Pauli operator at lattice site $j$ can be related to the field operators through the bosonization relations \cite{TG04},
\eqn{
\hat{\sigma}_{j}^{z}	&\simeq&\frac{2a}{\pi}\partial_{x}\hat{\phi}+c_{1}\left(-1\right)^{j}\cos\bigl(2\hat{\phi}\bigr),\label{sigmazb}\\
\hat{\sigma}_{j}^{+}	&\simeq& e^{i\hat{\theta}}\left[c_{2}\left(-1\right)^{j}+c_{3}\cos\bigl(2\hat{\phi}\bigr)\right],\label{sigmapb}
} where $a$ is the lattice spacing, and $c_{1,2,3}$ are nonuniversal coefficients. We here note that the spin quantization axis is chosen such that the total $z$ magnetization $S_{\rm tot}^z=\sum_{j}\hat{\sigma}^{z}_{j}/2$ corresponds to the conserved charge of the TLL.
The Euclidean action can be expressed in terms of either $\phi$ or $\theta$ representations by
\eqn{
{\cal S}_{0}[\phi]&=&\int dxd\tau\frac{1}{2\pi K}\left[\left(\partial_{x}\phi\right)^{2}+\left(\partial_{\tau}\phi\right)^{2}\right],\\
{\cal S}_{0}[\theta]&=&\int dxd\tau\frac{K}{2\pi}\left[\left(\partial_{x}\theta\right)^{2}+\left(\partial_{\tau}\theta\right)^{2}\right],
}
where each field is compactified on a circle as 
\eqn{\label{compcon}
\phi\sim\phi+\pi n,\;\;\theta\sim\theta+2\pi m,\;\;n,m\in\mathbb{Z}.
}The condition on $\phi$ reflects the quantization of the total magnetization, $S_{\rm tot}^z\in\mathbb{Z}$, while the condition on $\theta$ can be inferred from the bosonized expression of $\hat{\sigma}_j^+$ in Eq.~\eqref{sigmapb}.

We consider a local measurement process defined by the following Kraus operators:
\eqn{\label{kraus}
\hat{K}_{0,j}=\cos\zeta\,\hat{I},\;\;\hat{K}_{\pm,j}=\sin\zeta\,\frac{1\pm\hat{\sigma}_{j}^{\alpha}}{2}
}
with $0<\zeta\leq\pi/2$ and $\alpha\in\{x,y,z\}$. At each site, this process corresponds to performing the site-resolved projection measurement along axis $\alpha$ with probability $\sin^2\zeta$ and doing nothing otherwise. The resulting CPTP map $\cal{E}$ in Eq.~\eqref{krausproduct} has the interpretation as the unconditioned evolution, which corresponds to taking the ensemble average over the measurement outcomes, i.e.,  throwing away all the information acquired by the measurements (see, e.g., Ref.~\cite{YAreview}).
Alternatively, the unconditioned evolution $\cal E$ can be also regarded as the finite-time evolution of the Markovian master equation ${\cal E}=e^{{\cal L}t}$, which is generated by 
\eqn{{\cal L}(\hat{\rho})=-\frac{1}{2}\sum_{j}(\hat{L}_{j}^{\dagger}\hat{L}_{j}\hat{\rho}+\hat{\rho}\hat{L}_{j}^{\dagger}\hat{L}_{j}-2\hat{L}_{j}\hat{\rho}\hat{L}_{j}^{\dagger}).\label{liouville}} 
Here, the jump operators are given by
\eqn{
\hat{L}_{j}=\sqrt{\gamma}\hat{\sigma}_{j}^{\alpha}\label{jump}}
with $\gamma$ being the measurement rate. This simply means that the nonunitary evolution ${\cal E}$ corresponds to local decoherence or dephasing along axis $\alpha$ due to the Markovian environment. 

From Eqs.~\eqref{rhoed0} and \eqref{rhoed}, the post-measurement density matrix in the doubled Hilbert space can be obtained as 
\eqn{\label{rhoed2}
|\rho_{{\cal E}}\rparen\!=\exp\Bigl\{-\mu\Bigl[\sum_{j}(1-\hat{\sigma}_{j}^{\alpha}\otimes\hat{\sigma}_{j}^{\alpha})\Bigr]\Bigr\} |\Psi_{0}\rangle\!\otimes\!|\Psi_{0}\rangle,\label{channel}
}which corresponds to Eq.~\eqref{rhoed} with $\hat{k}_{j1}=\hat{\tilde{k}}_{j1}=\hat{I}$ and $\hat{k}_{j2}=\hat{\tilde{k}}_{j2}=i\hat{\sigma}_j^\alpha$.
Here, the measurement strength $\mu>0$ can be related to the parameters in Eqs.~\eqref{kraus} and \eqref{jump} via $\mu=-\log\cos\zeta$ and $\mu=\gamma t$, respectively.
Thus, the strong coupling limit $\mu\to\infty$ corresponds to the limit $\zeta\to \pi/2$ of performing the projection measurement at all the sites \cite{SJM09} or, equivalently, the long-time limit $t\to\infty$ of the Markovian evolution induced by Eq.~\eqref{liouville}. We note that in this limit the coherence is completely lost, and the density matrix reduces to a classical diagonal ensemble. 
Below we shall refer to the nonunitary evolution~\eqref{channel} along the symmetry axis $\alpha=z$ as density measurement and that on the easy plane $\alpha=x,y$ as phase measurement. This is because the spin-$\frac{1}{2}$ operator $\hat{\sigma}^z$ ($\hat{\sigma}^{x,y}$) has the interpretation as density (phase) fluctuations of 1D interacting particles as inferred from Eq.~\eqref{sigmazb}  (Eq.~\eqref{sigmapb}).

\section{Renormalization group analysis\label{sec:rg}}

Our main goal is to demonstrate that the TLL influenced by a local measurement~\eqref{channel} can exhibit entanglement phase transitions in the system-environment Hilbert space. In the field-theoretical formalism, these transitions correspond to the boundary phase transitions of the doubled fields in Eq.~\eqref{stotact}. To this end, we first need to perform a RG analysis to assess whether or not the boundary perturbation is relevant and determine which conformal boundary conditions are imposed in the IR limit.

\subsection{Density measurement\label{subsec:rgz}}
We first consider the case of density measurement in which decoherence along the symmetry axis $\alpha=z$ occurs. Using Eq.~\eqref{channel} and the bosonized expression~\eqref{sigmazb}, we can obtain the boundary action~\eqref{seact} as \footnote{Equation~\eqref{bactgz} can be inferred from the following  relation $\sum_{j}\left(1-\hat{\sigma}_{j}^{z}\otimes\hat{\sigma}_{j}^{z}\right)=\sum_{j}\frac{\left(1\otimes\hat{\sigma}_{j}^{z}-\hat{\sigma}_{j}^{z}\otimes1\right)^{2}}{2}
\simeq\!\!\int\!\frac{dx}{2a}[\frac{4a^{2}}{\pi^{2}}(\partial_{x}\hat{\phi}\!-\!\partial_{x}\hat{\tilde{\phi}})^{2}\!+\!c_{1}^{2}(\cos(2\hat{\phi})\!-\!\cos(2\hat{\tilde{\phi}}))^{2}]$.
}
\eqn{\label{bactgz}
{\cal S}_{{\cal E}}[\phi,\tilde{\phi}]&=&\mu\int dxd\tau\,\delta(\tau)\Bigl[\frac{2a}{\pi^{2}}(\partial_{x}\phi-\partial_{x}\tilde{\phi})^{2}\nonumber\\
&&+\frac{c_{1}^{2}}{2a}(\cos(2\phi)-\cos(2\tilde{\phi}))^{2}\Bigr],
}
where we use the path-integral formalism in the $\phi$ representation, and ${\phi}$ and ${\tilde{\phi}}$ are the two copies of the scalar fields. 

It is useful to introduce the symmetric and antisymmetric combinations of the bosonic fields by
\eqn{
\phi_{+}=2(\phi+\tilde{\phi}),\;\;\phi_{-}=2(\phi-\tilde{\phi}).
}  
As a result, the total action~\eqref{stotact} can be rewritten as
\eqn{
{\cal S}_{{\rm tot}}[\phi_{+},\phi_{-}]={\cal S}_{0}[\phi_{+}]+{\cal S}_{0}[\phi_{-}]+{\cal S}_{{\cal E}}[\phi_{+},\phi_{-}],
}where the bulk action is a $c=1$ CFT,
\eqn{
{\cal S}_{0}[\phi_{\pm}]=\int dxd\tau\frac{1}{16\pi K}\left[\left(\partial_{x}\phi_{\pm}\right)^{2}+\left(\partial_{\tau}\phi_{\pm}\right)^{2}\right].\label{s0}
}The boundary term acting on the $\tau=0$ line is given by
\eqn{
{\cal S}_{{\cal E}}[\phi_+,\phi_-]&=&\mu\int dxd\tau\,\delta(\tau)\Bigl[\frac{a}{2\pi^{2}}\left(\partial_{x}\phi_{-}\right)^{2}\nonumber\\
&+&\frac{c_{1}^{2}}{2a}\left(1-\cos\left(\phi_{+}\right)\right)\left(1-\cos\left(\phi_{-}\right)\right)\Bigr],\label{sez}
}which is nonnegative and satisfies ${\cal S}_{{\cal E}}[\phi_{+},\phi_{-}\!=\!0]=0$ as discussed before in Sec.~\ref{subsec:path}.
The nonunitary evolution thus acts as the boundary interaction that tends to lock the phase difference $\phi_-$. Physically, this phase locking has an interpretation as  wavefunction collapse due to  measurement \cite{GC07}. To see this, one can unravel the CPTP map $\cal E$ into an individual quantum trajectory, that is, a stochastic nonunitary evolution conditioned on the measurement outcomes (cf. Eq.~\eqref{krausproduct}). There, the Kraus operators~\eqref{kraus} effectively act as a quantum nondemolition measurement of $\phi$ operators. As such, quantum jumps in each trajectory tend to stochastically localize the many-body wavefunction represented in the $\phi$ basis \cite{YA15}. Such wavefunction collapse results in the suppression of off-diagonal elements in the density matrix, which is captured by the locking of $\phi_-$.

After integrating out the bulk parts, we can obtain the (1+0)-dimensional action for boundary degrees of freedom.  Specifically, we express the $\tau=0$ components and their Fourier transforms by
\eqn{
\varphi_{\pm}(x)\equiv\phi_{\pm}(x,\tau=0),\;\;\varphi_{k\pm}=\int dx\,\varphi_{\pm}(x)e^{ikx},
}respectively, and integrate out the $\tau\neq 0$ components by performing the Gaussian integrations. The resulting action is  
\eqn{\label{Sdephasing}
{\cal S}&=&\frac{1}{2}\int_{-\Lambda_{0}}^{\Lambda_{0}}\frac{dk}{2\pi}\left[\frac{|k|}{4\pi K}|\varphi_{k+}|^{2}+\left(\frac{|k|}{4\pi K}+\frac{\gamma k^{2}}{\Lambda_{0}}\right)|\varphi_{k-}|^{2}\right]\nonumber\\
&+&u\Lambda_{0}\int_{-\infty}^{\infty}dx\left(1-\cos\left(\varphi_{+}(x)\right)\right)\left(1-\cos\left(\varphi_{-}(x)\right)\right),\nonumber\\
}where $\Lambda_{0}=2\pi/a$ is the UV momentum cutoff and we introduce the dimensionless parameters by
\eqn{\gamma=\frac{2\mu}{\pi},\;\;u=\frac{\mu c_{1}^{2}}{4\pi}.\label{gammu}}
The scaling dimensions of perturbations in Eq.~\eqref{Sdephasing} are
\eqn{
\text{dim} \left[(\partial_x \varphi_-)^2\right]=2,\;\;\text{dim}\left[\cos(\varphi_{\pm})\right]=2K.\nonumber\\
}
When $K>1/2$, one may argue from the perturbative RG analysis that all the perturbations in Eq.~\eqref{Sdephasing}, which are proportional to $\gamma$ or $u$,  are irrelevant, and this should lead to the trivial value $g_{\cal E}/g_{\cal I}=1$. Such prediction, however, is at odds with Ref.~\cite{SJM09} that has found a nonzero value of $s_0$ in the limit of the projection measurement $\mu\to\infty$ even when $K>1/2$. This fact indicates that we must carefully analyze the action $\cal S$ by going beyond a perturbative treatment.

For this purpose, we employ a nonperturbative approach known as the functional RG (fRG). While we present technical details in Appendix~\ref{app:frg}, here we summarize the key points. First of all, we neglect the cross coupling $\cos(\varphi_+)\cos(\varphi_-)$ in the action as it has the scaling dimension $4K$ and should be less relevant compared to the other potential terms. As a result, the action can be decoupled into the two sectors that include either $\varphi_+$ or $\varphi_-$. On the one hand, the $+$ sector is equivalent to the action discussed in the earlier studies of quantum impurity problems~\cite{Kane92,Furusaki93}. In this case, one can make the duality argument in the strong corrugation limit, and it is well-established that the potential denoted by $\cos(\varphi_+)$ is relevant (irrelevant) when $K<1/2$ ($K>1/2$) at any $u_+$.  

\begin{figure}[b]
\includegraphics[width=80mm]{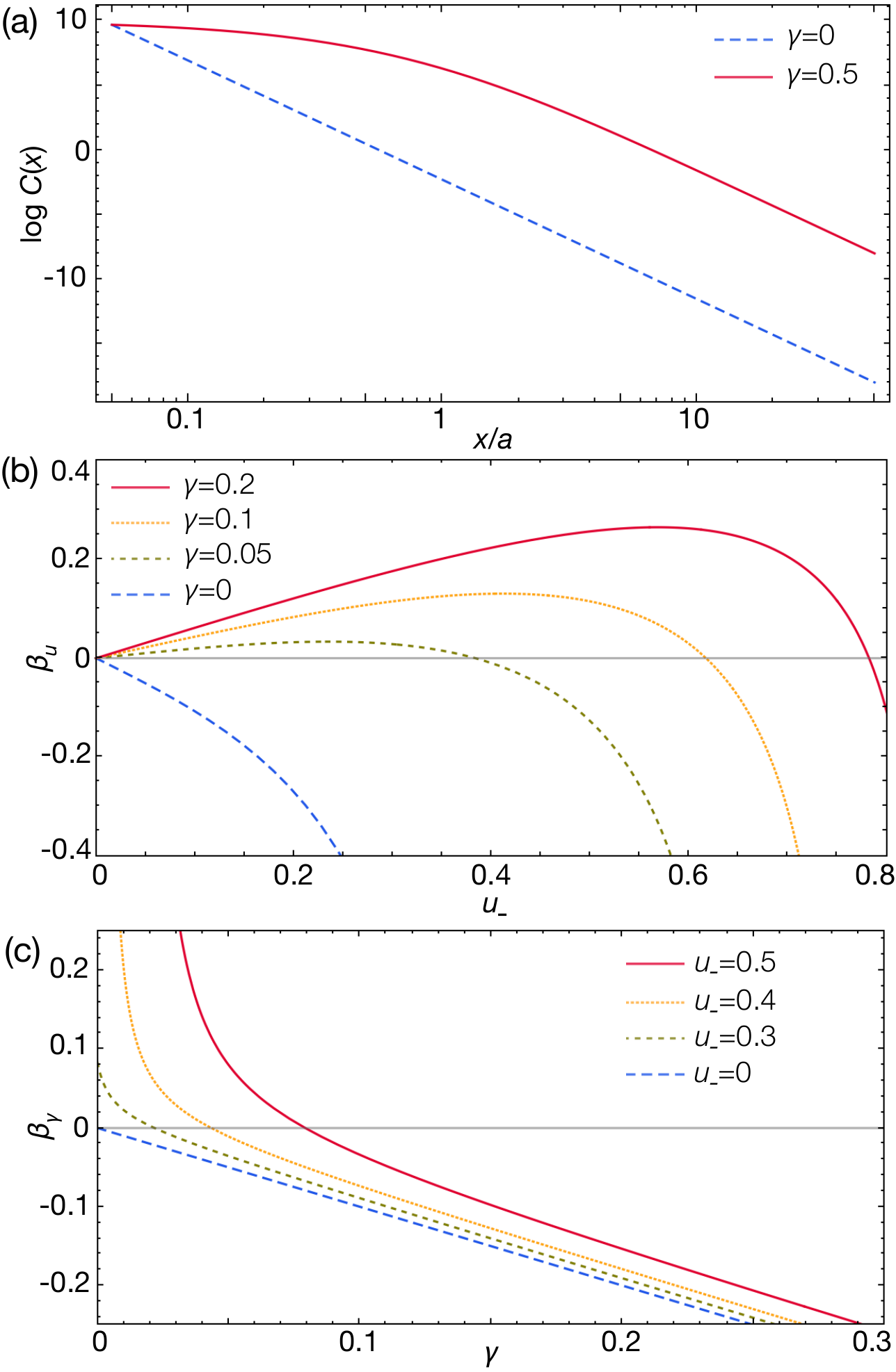} 
\caption{\label{fig_beta}
(a) Correlation function $C(x)$ plotted at the TLL parameter $K=1$ and the potential depth $u=0$. We set the normalization constants such that the correlation functions with different $\gamma$ take the same value at a UV scale $x/a=0.05$. (b,c) 
Beta functions $\beta_{u}$ in (b) and $\beta_{\gamma}$ in (c) plotted at $K=1$, where $u_-$ denotes the potential depth in the $-$ sector. Both of them are negatively valued in the perturbative limit $\gamma,u_{-}\to 0$, while they can acquire positive contributions in nonpertubrative regimes. 
}
\end{figure}

On the other hand, the $-$ sector requires a careful analysis because of the $k^2$ kinetic term in Eq.~\eqref{Sdephasing}, which is proportional to $\gamma$. 
It has been found in the context of the resistively shunted Josephson junction that the $\gamma$ term is dangerously irrelevant in the sense that it can be relevant in nonperturbative regions despite being perturbatively irrelevant \cite{MK22}. As demonstrated below, such anomalous enhancement of $\gamma$ can lead to the grow of the potential denoted by $\cos(\varphi_-)$ even when $K>1/2$, for which the potential is perturbatively irrelevant.  Intuitively, this point can be understood by observing that adding the $k^2$ kinetic term effectively decreases the value of $K$ for $\phi_-$ close to the boundary as inferred from Eq.~\eqref{Sdephasing}.  As such, one can expect that the $\gamma$ term effectively makes the boundary state susceptible to the cosine potential. To show this more explicitly, in Fig.~\ref{fig_beta}(a) we plot the crossover behavior of the correlation function $C(x)=\langle\cos(\varphi_-(x))\cos(\varphi_-(0))\rangle$, whose analytical expression is provided in Appendix~\ref{app:frg}. Its slow decay up to the crossover scale $x_c/a\sim 2\gamma K$ indicates that the scaling dimension of $\cos(\varphi_-)$ is indeed effectively close to zero at short distances. 

\begin{figure}[b]
\includegraphics[width=80mm]{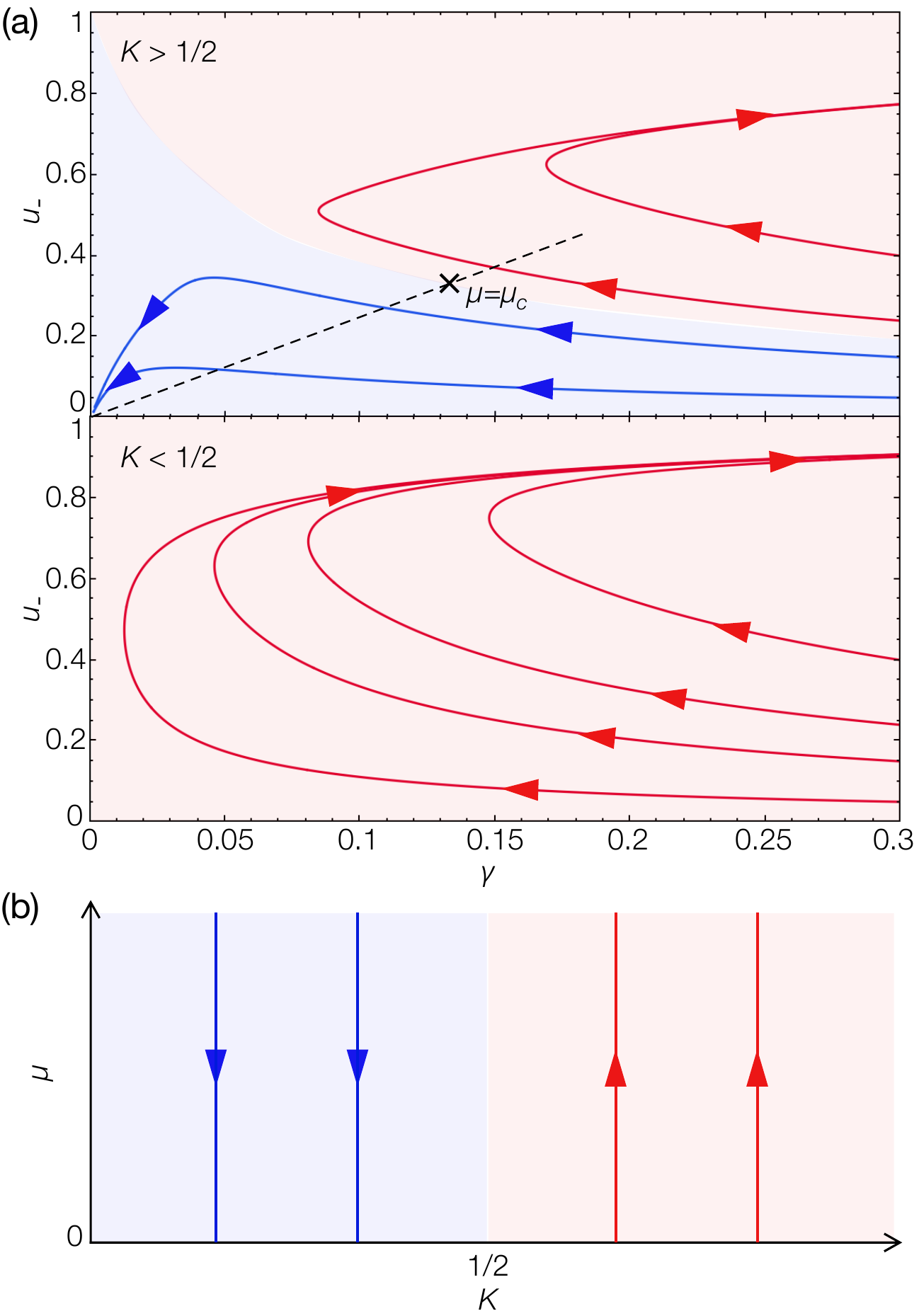} 
\caption{\label{fig_rg}
RG flow diagrams of the TLL under density measurement in (a) and phase measurement in (b) at different values of the TLL parameter $K$. In (a), the dashed line corresponds to varying the measurement strength $\mu$, and the crossing point indicates the critical value $\mu_c$ at which the transition occurs.
}
\end{figure}

We can derive the nonperturbative RG flow equations by
\eqn{\label{betafunc}
\frac{du_{-}}{dl}=\beta_{u}(u_{-},\gamma),\;\;\frac{d\gamma}{dl}=\beta_{\gamma}(u_{-},\gamma),
}where $l=\ln(\Lambda_0/\Lambda)$ is the logarithmic RG scale, $u_-$ denotes the depth of the potential $\cos(\varphi_-)$, and $\beta_{u,\gamma}$ are the beta functions whose full expressions are given in Appendix~\ref{app:frg}. The initial conditions at a UV scale $\Lambda=\Lambda_0$ are set by Eq.~\eqref{gammu}. In the perturbative limit $\gamma,u_-\ll 1$, we have the asymptotes, $\beta_{u}\simeq (1-2K)u_{-}$ and $\beta_{\gamma}\simeq -\gamma$, which are consistent with the scaling dimensions. In nonperturbative regions, however, both $\beta_{u,\gamma}$ can be positive as shown in Fig.~\ref{fig_beta}(b,c).  
The resulting RG flow diagrams are shown in Fig.~\ref{fig_rg}(a). Notably, when $K>1/2$, there is a critical value $\mu_c$ in the measurement strength $\mu$, above which the dangerously irrelevant term $\gamma$ leads to the nonmonotonic RG flows toward the strong coupling limit (top panel). This transition manifests itself as a singular change of the universal contribution $s_0$ in the system-environment entanglement $S_{SE}$ as we demonstrate later both analytically and numerically.

To determine the value of $s_0$ or, equivalently, the $g$ function, we need to identify conformal boundary conditions that are realized in the IR limit of boundary RG flows. In the present case, the grow of a potential term $u_{\pm}$ leads to the phase locking $\varphi_{\pm}=0$, which acts as the Dirichlet boundary condition (D.b.c.) of $\phi_{\pm}$, respectively.  
Accordingly, when $K>1/2$, there must exist a threshold value $\mu=\mu_c$ below which 
the boundaries of both $\phi_{\pm}$ remain free, i.e., obey the Neumann boundary condition (N.b.c.), and above which the boundary condition for only $\phi_-$ changes to the D.b.c. Meanwhile, when $K<1/2$, the D.b.c.'s should be imposed on both of $\phi_{\pm}$ at any $\mu>0$, which means that an arbitrarily weak coupling to the environment can generate a nontrivial contribution to the system-environment entanglement. 
Corresponding to these boundary conditions, we obtain the value of the $g$ function as follows:
\eqn{\label{gfuncdeph}
\frac{g_{{\cal E}}}{g_{{\cal I}}}=\begin{cases}
2K & \forall\mu>0,\,K<1/2\\
1 & \mu<\mu_{c},\,K>1/2\\
\sqrt{2K} & \mu>\mu_{c},\,K>1/2
\end{cases},
}
whose derivations by boundary CFT will be given in Sec.~\ref{sec:bcft}. Additionally, in Sec.~\ref{subsec:numz} we will numerically verify these results by the exact diagonalization of the spin-$\frac{1}{2}$ XXZ chain. This agreement between the field-theoretical and numerical results would also serve as a further support for the validity of the fRG analysis for the resistively shunted Josephson junction, which is described by the same effective field theory.

Interestingly, since the ratio $g_{\cal E}/g_{\cal I}$ in Eq.~\eqref{gfuncdeph} can exceed unity, the present system might appear to violate the $g$-theorem \cite{AI91,FD04,CH16}. We speculate that this unconventional behavior originates from the nonmonotonic boundary RG flows predicted by our nonperturbative analysis (see the top panel of Fig.~\ref{fig_rg}(a)). 
Namely, it is likely that the theory reached in the UV limit, which is the source of the flows represented by the red curves, will be given by the boundary fixed point at $\gamma\to\infty$ and $u_-\to 0$. 
Indeed, in the limit $\gamma^{-1},u_-\ll 1$, we have $\beta_{u}\simeq u_-$ and $\beta_{\gamma}\simeq-\gamma$, which implies the asymptote $u_-\propto \gamma^{-1}$ near the UV theory. 
 The diverging $\gamma$ should favor the D.b.c., $\varphi_-=\varphi_0$, while its localization position $\varphi_0$ remains undetermined due to the vanishing potential terms. As such, there formally exist infinitely many possible boundary states, which we expect to lead to the diverging $g$ function. If so, we may argue that the $g$-theorem still remains valid in the RG flows of Fig.~\ref{fig_rg}(a) in the sense that the $g$ function monotonically decreases from the infinite value to some finite constant.

Before closing this section, let us comment on the value of $K$ in Eq.~\eqref{gfuncdeph}. Since the critical ground state of a standard spin-$\frac{1}{2}$ XXZ chain corresponds to the TLL having $K\ge1/2$, one might wonder how the result in Eq.~\eqref{gfuncdeph} for $K<1/2$ can be tested in actual spin systems. Indeed, when $K<1/2$, a cosine potential in the bulk action is expected to be relevant, leading to doubly degenerate ground states associated with the translational symmetry breaking. If $|\Psi_0\rangle$ is chosen to be the translationally symmetric ground state of a finite-size system, the ``cat state"-like feature of this state leads to a positive contribution $-s_0=\log 2$ to the system-environment entanglement  \cite{SJM09}. A possible way to avoid this and realize the TLL having $K<1/2$ is to consider, for instance, a spin-$\frac{1}{2}$ chain at the transitions between the N{\'e}el and dimer ordered states, where the bulk cosine term disappears \cite{FH80,NK94,FS10}.

\subsection{Phase measurement\label{subsec:rgx}}
We next discuss the case of the TLL subject to phase measurement. We choose $\alpha=x$ in Eq.~\eqref{channel} without loss of generality. To begin with, we use the bosonization formula~\eqref{sigmapb} to express the boundary action by \footnote{Equation~\eqref{dephbact} can be obtained from the following bosonization relation $\sum_{j}\left(1-\hat{\sigma}_{j}^{x}\otimes\hat{\sigma}_{j}^{x}\right)=\sum_{j}\frac{\left(1\otimes\hat{\sigma}_{j}^{x}-\hat{\sigma}_{j}^{x}\otimes1\right)^{2}}{2}
\simeq\int\frac{dx}{a}2c_{2}^{2}(\cos(\hat{\theta})-\cos(\hat{\tilde{\theta}}))^{2}$.}
\eqn{\label{dephbact}
{\cal S}_{{\cal E}}[\theta,\tilde{\theta}]=\mu\int dxd\tau\,\delta(\tau)\frac{2c_{2}^{2}}{a}(\cos({\theta})-\cos({\tilde{\theta}}))^{2}.
}
We then follow the same procedure as before, but in the $\theta$ representation this time. Specifically, we introduce the symmetric and antisymmetric combinations of $\theta$ and $\tilde{\theta}$ as
\eqn{
\theta_{+}=\theta+\tilde{\theta},\;\;\theta_{-}=\theta-\tilde{\theta},
}
leading to the total action
\eqn{
{\cal S}_{{\rm tot}}[\theta_{+},\theta_{-}]={\cal S}_{0}[\theta_{+}]+{\cal S}_{0}[\theta_{-}]+{\cal S}_{{\cal E}}[\theta_{+},\theta_{-}].
}
Here, the bulk action is given by
\eqn{
{\cal S}_{0}[\theta_{\pm}]=\int dxd\tau\frac{K}{4\pi}\left[\left(\partial_{x}\theta_{\pm}\right)^{2}+\left(\partial_{\tau}\theta_{\pm}\right)^{2}\right],
}
and the boundary action at $\tau=0$ is
\eqn{
{\cal S}_{{\cal E}}[\theta_+,\theta_-]\!=\!\!\mu\!\!\int dxd\tau\delta(\tau)\frac{2c_{2}^{2}}{a}\left(1\!-\!\cos\left(\theta_{+}\right)\right)\!\left(1\!-\!\cos\left(\theta_{-}\right)\right).\nonumber\\
}
After integrating out the bulk degrees of freedom, we obtain the effective action as 
\eqn{\label{Sbitflip}
{\cal S}&=&\sum_{s=\pm}\frac{1}{2}\int_{-\Lambda_{0}}^{\Lambda_{0}}\frac{dk}{2\pi}\frac{K|k|}{\pi}|\vartheta_{ks}|^{2}\nonumber\\
&&+w\Lambda_{0}\int dx\left(1-\cos\left(\vartheta_{+}\right)\right)\left(1-\cos\left(\vartheta_{-}\right)\right),
}where $w=\frac{\mu c_{2}^{2}}{\pi}$ and $\vartheta_{\pm}$ are the boundary components defined by  
\eqn{
\vartheta_{\pm}(x)\equiv\theta_{\pm}(x,\tau=0),\;\;\vartheta_{k\pm}=\int dx\,\vartheta_{\pm}(x)e^{ikx}.
}

Again, we may neglect the cross coupling $\cos(\vartheta_+)\cos(\vartheta_-)$ which is less relevant than the leading terms $w_{\pm}\cos(\vartheta_{\pm})$ according to the scaling dimensions. We can then decompose the action into the two sectors that include either $\vartheta_+$ or $\vartheta_-$. A key difference from the case of density measurement  above is that here the actions in both sectors are completely equivalent due to the absence of the $k^2$ kinetic term (compare Eq.~\eqref{Sdephasing} with Eq.~\eqref{Sbitflip}). As such, the flow equations of the couplings can be simply written as
\eqn{
\frac{dw_\pm}{dl}=\left(1-\frac{1}{2K}\right)w_\pm,
}which indicate that $w_\pm$ are relevant (irrelevant) when $K>1/2$ ($K<1/2$). The lack of the dangerously irrelevant term $\gamma$ also allows us to recover the duality argument \cite{Kane92}, from which no intermediate fixed points are expected during the RG flows (cf. Fig.~\ref{fig_rg}(b)). Consequently, both fields $\vartheta_+$ and $\vartheta_-$ should obey the D.b.c.'s (N.b.c.'s) at an arbitrarily weak $\mu$ when $K>1/2$ ($K<1/2$). The value of the corresponding $g$ function is given by
\eqn{\frac{g_{{\cal E}}}{g_{{\cal I}}}=\begin{cases}
1 & \forall\mu>0,\,K<1/2\\
\frac{1}{2K} & \forall\mu>0,\,K>1/2
\end{cases},\label{gfuncbit}}which will be derived in the next section by using the boundary CFT techniques. We note that Eq.~\eqref{gfuncbit} is consistent with the $g$-theorem.

\section{Boundary CFT analysis\label{sec:bcft}}
The boundary perturbations discussed above should lead to certain conformal boundary conditions in the IR limit. 
On the one hand, the conformal boundary conditions for minimal models, such as the $c=1/2$ Ising CFT, have been well-understood owing to the finiteness of the conformal towers \cite{REB00}. Also, in a single-component TLL corresponding to a $c=1$ CFT, the Dirichlet and Neumann boundary conditions are believed to be the only possible conformal  boundary conditions. On the other hand, there is a large variety of possible boundary conditions in the case of multicomponent TLLs described by a $c>1$ CFT, and their full understanding still remains open.    
Accordingly, to derive the $g$ function of the present model, we need a careful treatment of conformal boundary states \cite{CJ04,JP88,CJ89,HO96,MO97,FS11,Affleck01,MO10,MO06,Hsu_2010}. In particular, it is necessary to identify the precise compactification conditions imposed on the multiple bosonic fields. Below we provide the construction of a class of conformal boundary states for multicomponent TLLs and use it to determine the universal contribution to the system-environment entanglement.  

\subsection{Conformal boundary states with mixed Dirichlet-Neumann boundary conditions\label{subsec:bcftm}}

\begin{figure}[t]
\includegraphics[width=87mm]{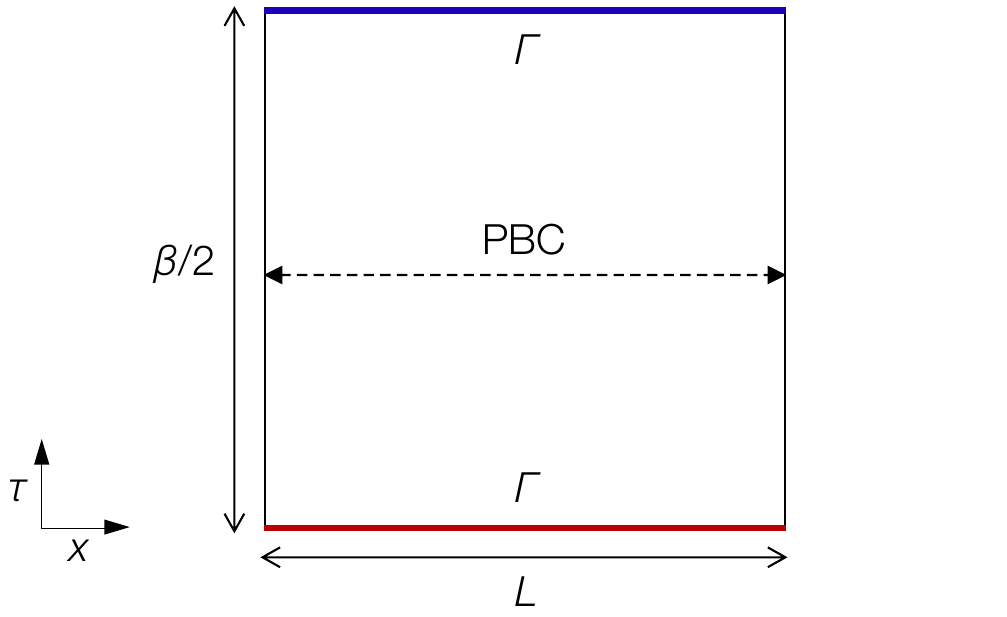} 
\caption{\label{fig_cft}
Conformal field theory of multicomponent free bosons~\eqref{freeb} is defined on the cylinder with circumference $L$ and length $\beta/2$. A boundary condition $\Gamma$ is imposed at both ends $\tau=0,\beta/2$. 
}
\end{figure}

We consider $N$-component bosonic fields $\boldsymbol{\Phi}$ governed by the Euclidean action
\eqn{\label{freeb}
{\cal S}_{0}[\boldsymbol{\Phi}]=\int\frac{dxd\tau}{2\pi}\left[\left(\partial_{x}\boldsymbol{\Phi}\right)^{2}+\left(\partial_{\tau}\boldsymbol{\Phi}\right)^{2}\right].
}As shown in Fig.~\ref{fig_cft}, the theory is defined on the (1+1)-dimensional sheets where the periodic boundary conditions are imposed on the spatial direction $x\in[0,L)$, while a certain boundary condition $\Gamma$ is imposed at both ends of the imaginary-time axis $\tau\in[0,\beta/2]$.  We aim to calculate the corresponding partition function by expressing it as the transition amplitude between the boundary states $|\Gamma\rangle$,
\eqn{\label{zgg}
Z_{\Gamma\Gamma}=\int{\cal D}\boldsymbol{\Phi}\,e^{-{\cal S}_{0}[\boldsymbol{\Phi}]}=\langle\Gamma|e^{-\frac{\beta}{2}\hat{H}_{{\rm CFT}}}|\Gamma\rangle.
}Here, $\hat{H}_{\rm CFT}$ is a Gaussian Hamiltonian of multicomponent fields
\eqn{\label{hcft0}
\hat{H}_{{\rm CFT}}=\int_{0}^{L}\frac{dx}{2\pi}\left[\left(\partial_{x}\hat{\boldsymbol{\Phi}}\right)^{2}+\left(\partial_{x}\hat{\boldsymbol{\Theta}}\right)^{2}\right],
}where the fields satisfy the commutation relation $[(\hat{\boldsymbol{\Phi}}(x))_{i},\partial_{x'}(\hat{\boldsymbol{\Theta}}(x'))_{j}]=i\pi\delta_{ij}\delta(x-x')$ and obey the periodic boundary conditions along the $x$ direction. We here assume that the bosonic fields $\boldsymbol{\Phi}$ are compactified as  
\eqn{
\boldsymbol{\Phi}&\sim&\boldsymbol{\Phi}+2\pi\boldsymbol{T},\;\;\boldsymbol{T}\in{\cal T},\\
{\cal T}&=&\Bigl\{ \boldsymbol{T}\,\Big{|}\,\boldsymbol{T}=\sum_{i=1}^{N}n_{i}\boldsymbol{a}_{i},n_{i}\in\mathbb{Z}\Bigr\},
}while the dual fields $\boldsymbol{\Theta}$ obey
\eqn{
\boldsymbol{\Theta}&\sim&\boldsymbol{\Theta}+2\pi\boldsymbol{T}^{*},\;\;\boldsymbol{T}^*\in{\cal T}^*/2,\\
{\cal T}^{*}/2&=&\Bigl\{ \boldsymbol{T}^{*}\,\Big{|}\,\boldsymbol{T}^{*}=\sum_{i=1}^{N}m_{i}\boldsymbol{b}_{i},m_{i}\in\mathbb{Z}\Bigr\},
}
where ${\cal T}^*$ is the reciprocal lattice of $\cal T$. The primitive vectors of $\cal T$ and ${\cal T}^*/2$ satisfy the relations $\boldsymbol{a}_{i}\cdot\boldsymbol{b}_{j}=\frac{1}{2}\delta_{ij}$.

We can expand these fields in terms of the oscillator modes and the zero modes generated by the windings along the spatial or temporal directions as follows:
\eqn{
\hat{\boldsymbol{\Phi}}\left(x,t\right)&=&\boldsymbol{\Phi}_{0}+\frac{2\pi}{L}\left(\hat{\boldsymbol{T}}x+\hat{\boldsymbol{T}}^{*}t\right)\nonumber\\
\!+\!&&\!\!\!\!\!\!\!\sum_{n=1}^{\infty}\!\!\frac{1}{\sqrt{4n}}\!\left[\hat{\boldsymbol{a}}_{n,{\rm L}}e^{-ik_{n}(x+t)}\!\!+\!\hat{\boldsymbol{a}}_{n,{\rm R}}e^{ik_{n}(x-t)}\!+\!{\rm H.c.}\right],\nonumber\\ \\
\hat{\boldsymbol{\Theta}}\left(x,t\right)&=&\boldsymbol{\Theta}_{0}+\frac{2\pi}{L}\left(\hat{\boldsymbol{T}}^{*}x+\hat{\boldsymbol{T}}t\right)\nonumber\\
\!+\!&&\!\!\!\!\!\!\!\sum_{n=1}^{\infty}\!\!\frac{1}{\sqrt{4n}}\!\left[\hat{\boldsymbol{a}}_{n,{\rm L}}e^{-ik_{n}(x+t)}\!\!-\!\hat{\boldsymbol{a}}_{n,{\rm R}}e^{ik_{n}(x-t)}\!+\!{\rm H.c.}\right],\nonumber\\
}where $\boldsymbol{\Phi}_{0}$ and $\boldsymbol{\Theta}_{0}$ are the  zero-mode angular variables, $\hat{\boldsymbol{T}}^*$ and $\hat{\boldsymbol{T}}$ are their conjugates,  $\hat{\boldsymbol{a}}_{n,{\rm L}({\rm R})}$ is a vector of annihilation operators of left- (right-) moving oscillator modes having quantum number $n$, and $k_{n}=2\pi n/L$. These operators satisfy the commutation relations
\eqn{
[(\boldsymbol{\Phi}_{0})_{i},(\hat{\boldsymbol{T}}^{*})_{j}]=\frac{i}{2}\delta_{ij},\;\;[(\boldsymbol{\Theta}_{0})_{i},(\hat{\boldsymbol{T}})_{j}]=\frac{i}{2}\delta_{ij},
}
\eqn{
[(\hat{\boldsymbol{a}}_{n,\alpha})_{i},(\hat{\boldsymbol{a}}_{m,\beta}^{\dagger})_{j}]=\delta_{nm}\delta_{\alpha\beta}\delta_{ij},
}where $i,j\in\{ 1,2,\ldots,N\}$, $\alpha,\beta\in\{ {\rm L},{\rm R}\}$, and $n,m\in\mathbb{N}$.  
Using these mode expansions, the Hamiltonian~\eqref{hcft0} can be expressed by
\eqn{\label{hcft}
\hat{H}_{{\rm CFT}}\!=\!\frac{2\pi}{L}\Bigl(\hat{\boldsymbol{T}}^{2}\!+\!\hat{\boldsymbol{T}}^{*2}\!+\!\sum_{n,\alpha,i}n\,\hat{a}_{n,\alpha,i}^{\dagger}\hat{a}_{n,\alpha,i}\!-\!\frac{N}{12}\Bigr),
}
where the last term originates from the Casimir energy due to the vacuum fluctuations of the oscillators.

We now suppose that the boundary condition $\Gamma$ is characterized by the condition that the fields obey the D.b.c.'s within certain subspace ${\cal V}_\Gamma$ in the $N$-dimensional vector space. Specifically, denoting the projection matrix onto ${\cal V}_\Gamma$ by $P_\Gamma$, the fields satisfy 
 \eqn{\label{vgamma}
P_\Gamma{\hat{\boldsymbol{\Phi}}}|\Gamma\rangle={\rm const.} \;\;\;\;\forall x\in[0,L),
 }
which implies
\eqn{\label{phibc}
P_\Gamma\,\partial_{x}\hat{\boldsymbol{\Phi}}|\Gamma\rangle=0\;\;\;\;\forall x\in[0,L).
}
We further assume that the remaining parts of $\boldsymbol{\Phi}$ obey the N.b.c.'s 
 \eqn{\label{vcgamma}
(1-P_\Gamma)\partial_t{\hat{\boldsymbol{\Phi}}}|\Gamma\rangle=0\;\;\;\;\forall x\in[0,L).
 }
 Using $\partial_{x}\hat{\boldsymbol{\Theta}}=\partial_{t}\hat{\boldsymbol{\Phi}}$, we can rewrite Eq.~\eqref{vcgamma} as the following constraint on the dual fields:
\eqn{\label{thetabc}
(1-P_\Gamma)\,\partial_{x}\hat{\boldsymbol{\Theta}}|\Gamma\rangle=0\;\;\;\forall x\in[0,L).
}

We aim to construct a conformal boundary state $|\Gamma\rangle$ that is consistent with the above boundary conditions~\eqref{phibc} and \eqref{thetabc}. To this end, we first introduce the Ishibashi states $\hat{S}_{\Gamma}|\boldsymbol{T},\boldsymbol{T}^{*}\rangle$, which consist of squeezed vacuum of  oscillator modes and have zero-mode quantum numbers $\boldsymbol{T}$ and $\boldsymbol{T}^*$. Here, the squeezing operator is defined by
\eqn{
\hat{S}_{\Gamma}=\exp\Bigl(-\sum_{n=1}^{\infty}\hat{\boldsymbol{a}}_{n,{\rm L}}^{\dagger}O\hat{\boldsymbol{a}}_{n,{\rm R}}^{\dagger}\Bigr)
}with $O$ being an orthogonal matrix, and the zero-mode states $|\boldsymbol{T},\boldsymbol{T}^{*}\rangle$ satisfy the relations
\eqn{
\hat{\boldsymbol{T}}|\boldsymbol{T},\boldsymbol{T}^{*}\rangle&=&\boldsymbol{T}|\boldsymbol{T},\boldsymbol{T}^{*}\rangle,\\
\hat{\boldsymbol{T}}^{*}|\boldsymbol{T},\boldsymbol{T}^{*}\rangle&=&\boldsymbol{T}^{*}|\boldsymbol{T},\boldsymbol{T}^{*}\rangle,\\
\hat{a}_{n,\alpha,i}|\boldsymbol{T},\boldsymbol{T}^{*}\rangle&=&0.
}
Thus, the states $\hat{S}_{\Gamma}|\boldsymbol{T},\boldsymbol{T}^{*}\rangle$ satisfy the boundary conditions \eqref{phibc} and \eqref{thetabc} by setting
 \eqn{
\boldsymbol{T}\in{\cal T}_{\Gamma}={\cal T}\cap{{\cal V}}^\perp_{\Gamma}
,&&\;\boldsymbol{T}^{*}\in{\cal T}_{\Gamma}^{*}={\cal T}^{*}/2\cap{\cal V}_{\Gamma},\label{complat}\\
\;O&=&2P_{\Gamma}-I,
 }
 where ${{\cal V}}^\perp_{\Gamma}$ is the orthogonal complement of ${{\cal V}}_{\Gamma}$.
 
This construction, however, is not enough to define physical conformal boundary states, which must satisfy both the conformal invariance and  Cardy's consistency condition. On the one hand, a sufficient condition to ensure the conformal invariance of a candidate boundary state $|\Gamma\rangle$ can be written as \cite{Affleck01}
\eqn{\label{cinv}
\left(\hat{\boldsymbol{\alpha}}_{n,{\rm L}}-O\hat{\boldsymbol{\alpha}}_{-n,{\rm R}}\right)|\Gamma\rangle=0\;\;\;\;\forall n\in\mathbb{Z}
}for some orthogonal matrix $O$ and the vectors
\eqn{
\hat{\boldsymbol{\alpha}}_{n,{\rm L}}\!=\!\!\begin{cases}
\hat{\boldsymbol{a}}_{n,{\rm L}} & n>0\\
\hat{\boldsymbol{T}}+\hat{\boldsymbol{T}}^{*} & n=0\\
-\hat{\boldsymbol{a}}_{-n,{\rm L}}^{\dagger} & n<0
\end{cases},\;\;
\hat{\boldsymbol{\alpha}}_{n,{\rm R}}\!=\!\!\begin{cases}
\hat{\boldsymbol{a}}_{n,{\rm R}} & n>0\\
-\hat{\boldsymbol{T}}+\hat{\boldsymbol{T}}^{*} & n=0\\
-\hat{\boldsymbol{a}}_{-n,{\rm R}}^{\dagger} & n<0
\end{cases}.\nonumber\\
}
One can readily check that the Ishibashi states $\hat{S}_{\Gamma}|\boldsymbol{T},\boldsymbol{T}^{*}\rangle$ satisfy the condition~\eqref{cinv}, meaning that they are conformally invariant. On the other hand, they do not satisfy the Cardy's condition that is imposed on the partition function after the modular transformation \cite{CJ89}. 
In fact, it is the linear combination of them that satisfies this condition and thus acts as a legitimate conformal boundary state \cite{Affleck01,MO06}:
\eqn{\label{gammastate}
|\Gamma\rangle=g_{\Gamma}\sum_{\boldsymbol{T}\in{\cal T}_{\Gamma}}\sum_{\boldsymbol{T}^{*}\in{\cal T}_{\Gamma}^{*}}\hat{S}_{\Gamma}|\boldsymbol{T},\boldsymbol{T}^{*}\rangle,
}where we introduce the coefficient $g_\Gamma$, which plays the role of the $g$ function as shown below.
While Eq.~\eqref{gammastate} defines only a subclass of boundary states among all the possible conformally invariant boundary states, this is enough for our purpose of identifying the $g$ function in the TLL under a local measurement.

To determine the value of $g_\Gamma$, we consider the modular transformation of the partition function. Namely, we use Eqs.~\eqref{zgg}, \eqref{hcft}, and \eqref{gammastate} to get
\eqn{
Z_{\Gamma\Gamma}=\frac{g_{\Gamma}^{2}}{\left(\eta\left(q\right)\right)^{N}}\sum_{\boldsymbol{T}\in{\cal T}_{\Gamma}}\sum_{\boldsymbol{T}^{*}\in{\cal T}_{\Gamma}^{*}}q^{\frac{1}{2}\left(\boldsymbol{T}^{2}+\boldsymbol{T}^{*2}\right)},
}where $q=e^{-2\pi\beta/L}$ and $\eta(q)=q^{\frac{1}{24}}\prod_{n=1}^{\infty}\left(1-q^{n}\right)$ is the Dedekind $\eta$ function \cite{RB09}. After performing the modular transformation, where the roles of space and time are exchanged, we can express the partition function by
\eqn{\label{zgg0}
Z_{\Gamma\Gamma}=\frac{g_{\Gamma}^{2}}{v_{0}\left({\cal T}_{\Gamma}\right)v_{0}\left({\cal T}_{\Gamma}^{*}\right)\left(\eta\left(\tilde{q}\right)\right)^{N}}\sum_{\tilde{\boldsymbol{T}}\in\tilde{{\cal T}}_{\Gamma}}\sum_{\tilde{\boldsymbol{T}}^{*}\in\tilde{{\cal T}}_{\Gamma}^{*}}\tilde{q}^{\frac{1}{2}\left(\tilde{\boldsymbol{T}}^{2}+\tilde{\boldsymbol{T}}^{*2}\right)}.\nonumber\\
}Here, we use the multidimensional Poisson formula to derive the right-hand side, $v_0(\cdot)$ denotes the unit-cell volume of the concerned compactification lattice, and $\tilde{q}=e^{-2\pi L/\beta}$. We recall that the unit-cell volume can be obtained as $v_0({\cal W})=\sqrt{\text{det}(W)}$, where $W_{ij}=\boldsymbol{s}_{i}\cdot\boldsymbol{s}_{j}$ is the Gram matrix of the primitive vectors $\boldsymbol{s}_i$ of a lattice $W$. The leading contribution to Eq.~\eqref{zgg0} in the limit $L\gg\beta$ is given by
\eqn{\label{zgg1}
Z_{\Gamma\Gamma}\simeq\frac{g_{\Gamma}^{2}}{v_{0}\left({\cal T}_{\Gamma}\right)v_{0}\left({\cal T}_{\Gamma}^{*}\right)}e^{\frac{\pi NL}{12\beta}}.
}Meanwhile, it should be also possible to interpret $Z_{\Gamma\Gamma}$ as the partition function of a 1D quantum system of finite length $\beta/2$ and the inverse temperature $L$ with boundary conditions $\Gamma$ at the two ends $\tau=0,\beta/2$ of the ``spatial axis" $\tau$. Namely, we should be able to express the partition function as 
\eqn{\label{zqexp}
Z_{\Gamma\Gamma}={\rm tr}\,e^{-L\hat{H}^{\Gamma\Gamma}_{\rm CFT}},} where $\hat{H}^{\Gamma\Gamma}_{\rm CFT}$ is the CFT Hamiltonian subject to the boundary conditions $\Gamma$ at both ends. 
When the ground state of $\hat{H}^{\Gamma\Gamma}_{\rm CFT}$ is unique, we get in the ``zero-temperature" limit $L\gg\beta$
\eqn{\label{zgg2}
Z_{\Gamma\Gamma}\simeq e^{\frac{\pi NL}{12\beta}},
}where we use the fact that the zero-point energy of a CFT Hamiltonian on a finite chain of length $l$ with edges is given by $-\pi c/(24l)$ \cite{RB09}.  Comparing the coefficients of Eqs.~\eqref{zgg1} and \eqref{zgg2}, we obtain the formula of the $g$ function by
\eqn{\label{gformula}
g_{\Gamma}=\sqrt{v_{0}\left({\cal T}_{\Gamma}\right)v_{0}\left({\cal T}_{\Gamma}^{*}\right)}.
}

We now explain the interpretation of $g_\Gamma$ as an effective ground-state degeneracy as follows \cite{AI91}. In the language of Eq.~\eqref{zqexp}, where the roles of space and time are exchanged, the above limit corresponds to taking the zero-temperature limit $1/L\to 0$ first, before taking the infinite-size limit $\beta\to\infty$. Since any finite-size quantum system has a discrete spectrum, the degeneracy of ground states must be integer valued in this case. 
The interpretation of the $g$ function as a noninteger ground-state degeneracy becomes evident if we take the limits in the opposite order. 
Namely, taking the infinite-size limit $\beta\to\infty$ first, a quantum Hamiltonian can exhibit a continuous spectrum, and the system can have a noninteger ``ground-state degeneracy". In the zero-temperature limit $1/L\to 0$, this degeneracy gives rise to a temperature-independent constant in thermal entropy. Such contribution is nothing but a $L$-independent multiplicative factor of the partition function which is introduced as the $g$ function in Eq.~\eqref{zxi}. We note that the term linear in $L$ in Eq.~\eqref{zxi} originates from the unregulated part of the partition function, which explicitly depends on the short-distance cutoff through the ratio $L/a$; as a result, this term is nonuniversal.

For the sake of later convenience, we discuss the case in which one imposes two different boundary conditions $\Gamma_{1,2}$ at $\tau=0,\beta/2$, respectively. The partition function can be represented by the amplitude between the two boundary states 
\eqn{Z_{\Gamma_{1}\Gamma_{2}}=\langle\Gamma_{1}|e^{-\frac{\beta}{2}\hat{H}_{{\rm CFT}}}|\Gamma_{2}\rangle.} 
Its leading contribution in the limit $\beta\gg L$ is then given by \eqn{\label{gdeg}
Z_{\Gamma_{1}\Gamma_{2}}\simeq\langle\Gamma_{1}|{\rm GS}\rangle\langle{\rm GS}|\Gamma_{2}\rangle e^{\frac{\pi N\beta}{12L}}=g_{\Gamma_{1}}g_{\Gamma_{2}}e^{\frac{\pi N\beta}{12L}},
}where we use Eqs.~\eqref{hcft}, \eqref{gammastate}, and the fact that the ground state of $\hat{H}_{\rm CFT}$ is a vacuum state having the vanishing zero-mode numbers $\boldsymbol{T}=\boldsymbol{T}^*=\boldsymbol{0}$, i.e., $|{\rm GS}\rangle=|\boldsymbol{0},\boldsymbol{0}\rangle$. 
One can see that the coefficient introduced in Eq.~\eqref{gammastate} indeed appears as the $L$-independent contribution to the partition function, which has an interpretation as the ground-state degeneracy in the sense explained above.

\begin{figure}[b]
\includegraphics[width=80mm]{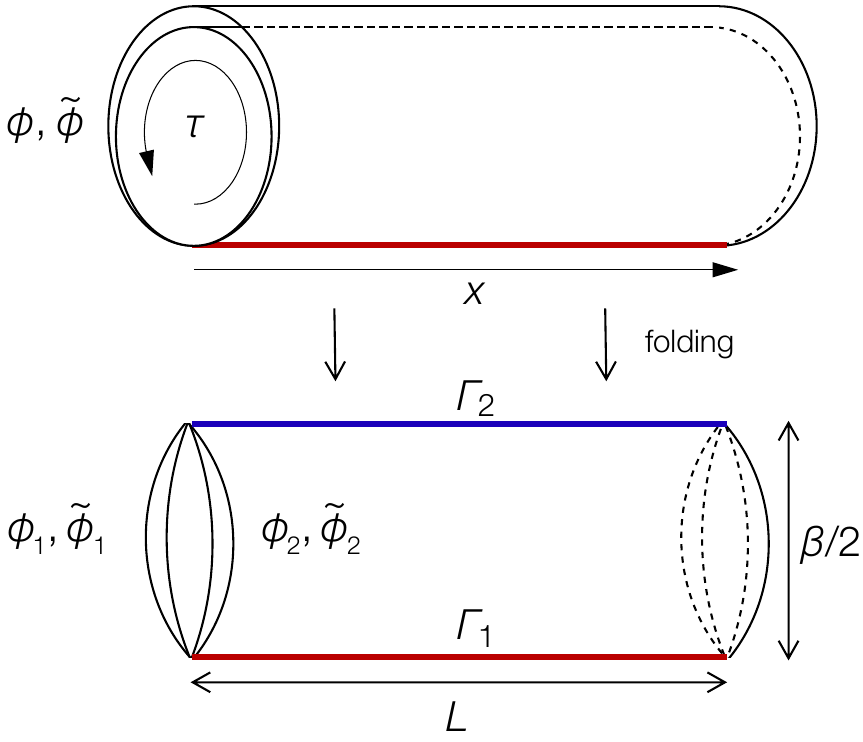} 
\caption{\label{fig_cft2}
The upper figure shows the two copies of the free bosonic fields $\phi,\tilde{\phi}$ defined on the torus of size $L\times \beta$. The periodic boundary condition is imposed on the spatial direction $x$. At low energies, a boundary interaction at $\tau=0$ gives rise to a certain conformal boundary condition as indicated by the red line. In the lower figure, the system is folded into the theory of the four-component fields $\phi_1,\tilde{\phi}_1,\phi_2,\tilde{\phi}_2$ defined on the cylinder with circumference $L$ and length $\beta/2$. Boundary conditions denoted by $\Gamma_1$ and $\Gamma_2$ are imposed at the ends $\tau=0,\beta/2$ as indicated by the red and blue lines, respectively.
}
\end{figure}

\subsection{Boundary CFT in the doubled Hilbert space\label{subsec:bcd}}
We now derive the $g$ function of the TLL under a local measurement by using the boundary CFT results above. To do so, we recall that in the doubled Hilbert space formalism the low-energy theory contains the two copies of the bosonic fields $\phi,\tilde{\phi}$ defined on the torus of size $L\times \beta$. A nonunitary evolution is then represented as the boundary interaction ${\cal S}_{\cal E}$ acting on the $\tau=0$ line. To calculate the universal contribution, it is useful to fold the system by doubling the number of fields as shown in Fig.~\ref{fig_cft2}, where the space-time geometry becomes topologically equivalent to a cylinder after the folding. In this way, we can map the problem to the theory of the four-component fields $\phi_1,\tilde{\phi}_1,\phi_2,\tilde{\phi}_2$ defined on the cylinder with circumference $L$ and length $\beta/2$. Boundary conditions satisfied at the ends $\tau=0,\beta/2$ of the cylinder are denoted by $\Gamma_{1,2}$, respectively.

To apply the boundary CFT results to the present system, we identify the fields in Eq.~\eqref{hcft0} as 
\eqn{
\boldsymbol{\Phi}&=&\frac{1}{\sqrt{K}}\left(\phi_{1},\phi_{2},\tilde{\phi}_{1},\tilde{\phi}_{2}\right)^{{\rm T}},\\
\boldsymbol{\Theta}&=&\sqrt{K}\left(\theta_{1},\theta_{2},\tilde{\theta}_{1},\tilde{\theta}_{2}\right)^{{\rm T}},
}
where the TLL parameter $K$ is included in the definitions. The original fields are compactified as (cf. Eq.~\eqref{compcon})
\eqn{
\phi_{i}&\sim&\phi_{i}+\pi n_i,
\;\;
\tilde{\phi}_{i}\sim\tilde{\phi}_{i}+\pi \tilde{n}_i,\\
\theta_{i}&\sim&\theta_{i}+2\pi m_i,\;\;
\tilde{\theta}_{i}\sim\tilde{\theta}_{i}+2\pi \tilde{m}_i,
}
where $i\in\{1,2\}$ and $n_i,\tilde{n}_i,m_i,\tilde{m}_i\in{\mathbb Z}$. 
The corresponding compactification conditions on the fields $\boldsymbol{\Phi}$ and $\boldsymbol{\Theta}$ are described by
\eqn{
\boldsymbol{\Phi}&\sim&\boldsymbol{\Phi}+2\pi\boldsymbol{T},\;\;\boldsymbol{T}\in{\cal T},\\
{\cal T}&=&\Bigl\{ \boldsymbol{T}\,\Big{|}\,\boldsymbol{T}=\sum_{l=1}^{4}n_{l}\boldsymbol{a}_{l},n_{l}\in\mathbb{Z},\boldsymbol{a}_{l}=\frac{1}{2\sqrt{K}}\boldsymbol{e}_{l}\Bigr\},
\label{tcom}
}
and
\eqn{
\boldsymbol{\Theta}&\sim&\boldsymbol{\Theta}+2\pi\boldsymbol{T}^*,\;\;\boldsymbol{T}^*\in{\cal T}^*/2,\\
{\cal T}^*/2&=&\Bigl\{ \boldsymbol{T}^*\,\Big{|}\,\boldsymbol{T}^*=\sum_{l=1}^{4}m_{l}\boldsymbol{b}_{l},m_{l}\in\mathbb{Z},\boldsymbol{b}_{l}=\sqrt{K}\boldsymbol{e}_{l}\Bigr\},\nonumber\\
\label{tscom}
}
respectively, where $\boldsymbol{e}_{l}$ with $l\in\{1,2,3,4\}$ are the unit vectors.
The $g$ function of the present multicomponent field theory, which is denoted by $g_{\cal E}$, is then given by the products of $g_{\Gamma_1}$ and $g_{\Gamma_2}$ that are defined as the coefficients in the boundary states $|\Gamma_{1,2}\rangle$ at each end (see Eq.~\eqref{gdeg}). When necessary, we also have to include an additional integer degeneracy $d$ that accounts for multiplicity of possible boundary conditions \cite{CV09}. Taken together, we have
\eqn{
g_{{\cal E}}=g_{\Gamma_{1}}g_{\Gamma_{2}}d.
}
Below we determine a value of the universal contribution~\eqref{univcont} to the system-environment entanglement on the basis of this formalism.

\subsection{Density measurement\label{subsec:bcftd}}
\subsubsection*{Calculation of $g_{\Gamma_2}$}
We  first discuss the case of the TLL under density measurement. We start by considering the $g$ function of the boundary $\Gamma_2$ at $\tau=\beta/2$ which is created by the folding procedure. 
Since there were merely two bosonic fields $\phi,\tilde{\phi}$ on the torus before the folding, the boundary conditions at $\Gamma_2$ are simply given by
\eqn{\label{bc0}
\phi_{1}=\phi_{2},\;\;\tilde{\phi}_{1}=\tilde{\phi}_{2}.
}
Accordingly, the subspace ${\cal V}_{\Gamma_2}$ defining the D.b.c.'s is 
\eqn{{\cal V}_{\Gamma_2}={\rm span}\left(\left(\begin{array}{c}
1\\
-1\\
0\\
0
\end{array}\right),\left(\begin{array}{c}
0\\
0\\
1\\
-1
\end{array}\right)\right).
}
From Eq.~\eqref{tscom}, the unit-cell volume of ${\cal T}^*_{\Gamma_2}={\cal T}^*/2\cap{\cal V}_{\Gamma_2}$ is then given by $|\boldsymbol{b}_1-\boldsymbol{b}_2|\cdot |\boldsymbol{b}_3-\boldsymbol{b}_4|=2K$. Meanwhile, the fields belonging to its orthogonal complement ${{\cal V}^\perp_{\Gamma_2}}$ remain free at the boundary, i.e., they obey the N.b.c.'s. From Eq.~\eqref{tcom}, the unit-cell volume of ${\cal T}_{\Gamma_2}={\cal T}\cap{{\cal V}^\perp_{\Gamma_2}}$ is $|\boldsymbol{a}_1+\boldsymbol{a}_2|\cdot |\boldsymbol{a}_3+\boldsymbol{a}_4|=1/(2K)$. Thus, the formula~\eqref{gformula} allows us to get
\eqn{
g_{\Gamma_2}=\sqrt{\underbrace{(2K)^{-1}}_{v_{0}({\cal T}_{\Gamma_{2}})}\times \underbrace{2K}_{v_{0}({\cal T}_{\Gamma_{2}}^{*})}}=1.
}
This result is consistent with the fact that boundary $\Gamma_2$ is nothing more than an artificial boundary created by the folding procedure, where no boundary entropy is expected.  

We note that the conditions~\eqref{bc0} must be satisfied also at the other edge $\Gamma_1$ because the fields obey the periodic boundary conditions along the imaginary-time $\tau$ axis.
In particular, when all the boundary interactions in ${\cal S}_{\cal E}$ are irrelevant and the fields obey only Eq.~\eqref{bc0}, the same result holds true as follows:
\eqn{
g_{\Gamma_1}=1,
}
which also implies $g_{\cal I}=1$. 
This simply means that in this case the system is a decoupled tori in the IR limit, which should not have any boundary entropy.
In contrast, when a boundary interaction is relevant, the edge $\Gamma_1$ is subject to a nontrivial boundary condition in addition to Eq.~\eqref{bc0}.
Our RG analyses suggest that there are several possible boundary conditions depending on  the measurement procedures and the values of $K$ and $\mu$, leading to distinct values of the $g$ function.

\begin{figure*}[t]
\includegraphics[width=179mm]{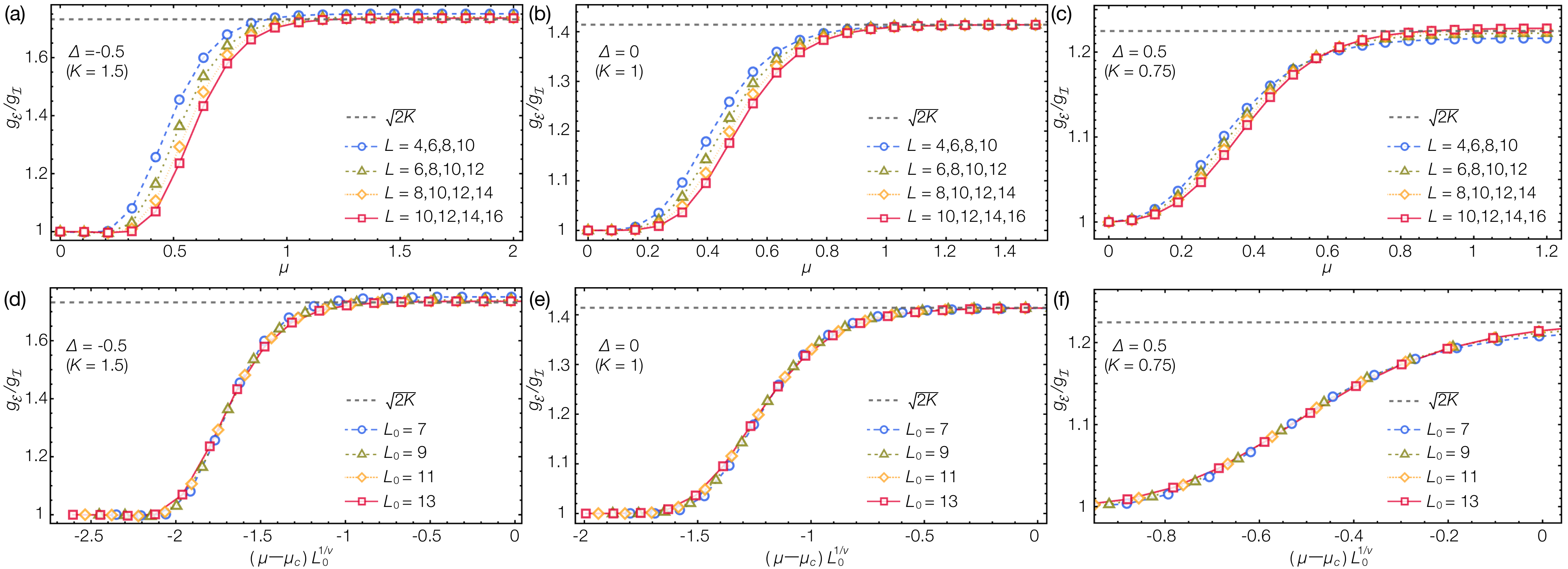} 
\caption{\label{fig_z_g}
Universal contribution $e^{s_0}=g_{\cal E}/g_{\cal I}$ to the system-environment entanglement in the TLL under density measurement. The numerical values are extracted from the second-order R\'enyi entropy fitted to $s_{1}L-s_{0}+\frac{s_{-1}}{L}$ with $L\in\{L_{0}-3,L_{0}-1,L_{0}+1,L_{0}+3\}$ and $L_0=7,9,11,13$. (a)-(c) Numerical results of the universal contribution at different system sizes and various $\Delta$. (d)-(f) Plots of the corresponding data collapses. The dashed horizontal lines in all the panels indicate the analytical values obtained by combining the boundary CFT results with Eq.~\eqref{bethe} that relates the TLL parameter $K$ to the anisotropy $\Delta$.
}
\end{figure*}

\subsubsection*{Case of $K>1/2$}
In the case of the $K>1/2$ TLL under density measurement, the RG analysis in Sec.~\ref{subsec:rgz} predicts that there is a threshold $\mu_c$ in the measurement strength $\mu$, below which the boundary perturbation is irrelevant, i.e., $g_{\cal E}=1$. Meanwhile, when $\mu>\mu_c$, the boundary interaction $\cos(\varphi_-)$ becomes relevant and localizes the field $\phi_-$ at $\tau=0$, which gives rise to the following boundary conditions imposed on $\Gamma_1$: 
\eqn{
\phi_{i}=\tilde{\phi}_{i}\;\; {\rm mod}\;\; \pi,\;\;i\in\{1,2\}.
}
Together with the boundary conditions~\eqref{bc0}, the corresponding Dirichlet subspace ${\cal V}_{\Gamma_1}$ is given by
\eqn{
{\cal V}_{\Gamma_{1}}={\rm span}\left(\left(\begin{array}{c}
1\\
-1\\
0\\
0
\end{array}\right),\left(\begin{array}{c}
0\\
0\\
1\\
-1
\end{array}\right),\left(\begin{array}{c}
1\\
0\\
-1\\
0
\end{array}\right)\right).
}
Accordingly, its orthogonal complement ${{\cal V}^\perp_{\Gamma_{1}}}$ is the one-dimensional vector space spanned by $(1,1,1,1)^{\rm T}$, in which the field obeys the N.b.c.'s. In a similar manner as described above, we can use Eqs.~\eqref{complat}, \eqref{gformula}, \eqref{tcom}, and \eqref{tscom}  to  obtain the $g$ function of the boundary state at $\Gamma_1$ as
\eqn{
g_{\Gamma_{1}}=\sqrt{\underbrace{K^{-1/2}}_{v_{0}({\cal T}_{\Gamma_{1}})}\times \underbrace{2K^{3/2}}_{v_{0}({\cal T}_{\Gamma_{1}}^{*})}}=\sqrt{2K}.
}
We recall that the boundary $\Gamma_2$ is an artificial boundary created by the folding and should have $g_{\Gamma_2}=1$. Since there is no additional degeneracy in the present case, we get
\eqn{\label{gs2k}
g_{{\cal E}}=\underbrace{\sqrt{2K}}_{g_{\Gamma_{1}}}\times\underbrace{1}_{g_{\Gamma_{2}}}\times\underbrace{1}_{d}=\sqrt{2K}.
}
We note that the value is consistent with the previous result obtained in the case of the projection measurement \cite{SJM09,MO10,Hsu_2010}.

\subsubsection*{Case of $K<1/2$}
We next consider the case of $K<1/2$ where the boundary interactions $\cos(\varphi_{\pm})$ in both sectors $\pm$ are relevant at any $\mu>0$. In this case, both of $\phi_{\pm}$ are locked at $\tau=0$, and we have the following boundary conditions at $\Gamma_1$:
\eqn{\label{d2bc}
\phi_{i}=\tilde{\phi}_{i}\;\;{\rm mod}\;\;\pi,\;\;\phi_{i}=-\tilde{\phi}_{i}\;\;{\rm mod}\;\;\pi,\;\;i\in\{1,2\}.
}
These conditions together with Eq.~\eqref{bc0} fully localize the field $\boldsymbol{\Phi}$. As such, the Dirichlet subspace ${\cal V}_{\Gamma_1}$ corresponds to the entire four-dimensional vector space in this case. Using Eqs.~\eqref{complat}, \eqref{gformula}, and \eqref{tscom}, we get
\eqn{
g_{\Gamma_{1}}=\sqrt{v_{0}\left({\cal T}^{*}/2\right)}=K.
}
Meanwhile, we note that the conditions~\eqref{bc0} and \eqref{d2bc} allow for the two possible D.b.c.'s associated with $\phi_{1}=\phi_{2}=\tilde{\phi}_{1}=\tilde{\phi}_{2}=0$ or $\phi_{1}=\phi_{2}=\tilde{\phi}_{1}=\tilde{\phi}_{2}=\frac{\pi}{2}$, which correspond to the degenerate potential minima in the boundary interactions. Including this additional two-fold degeneracy $d=2$, we have
\eqn{
g_{{\cal E}}=\underbrace{K}_{g_{\Gamma_{1}}}\times\underbrace{1}_{g_{\Gamma_{2}}}\times\underbrace{2}_{d}=2K.
}
The results are summarized in Eq.~\eqref{gfuncdeph}.

\subsection{Phase measurement\label{subsec:bcftb}}
We next discuss the $g$ function of the $K>1/2$ TLL under phase measurement. As described in Sec.~\ref{subsec:rgx}, the perturbative RG analysis together with the duality argument suggests that the boundary interactions $\cos(\vartheta_{\pm})$ in both sectors $\pm$ are relevant at $\forall \mu>0$ in this case. The relevant boundary perturbations thus lead to the following constraints at $\Gamma_1$: 
\eqn{\label{thebc1}
\theta_{i}=\tilde{\theta}_{i}\;\;{\rm mod}\;\;2\pi,\;\;\theta_{i}=-\tilde{\theta}_{i}\;\;{\rm mod}\;\;2\pi,\;\;i\in\{1,2\},
}
in addition to the periodic boundary conditions
\eqn{\label{thebc2}
\theta_{1}=\theta_{2},\;\;\tilde{\theta}_{1}=\tilde{\theta}_{2}.
}
The conditions~\eqref{thebc1} and \eqref{thebc2} fully localize the field $\boldsymbol{\Theta}$. Said differently, its dual $\boldsymbol{\Phi}$ obeys the N.b.c.'s because of the relation $\partial_{x}\boldsymbol{\Theta}=\partial_{\tau}\boldsymbol{\Phi}$, which means that the orthogonal complement ${{\cal V}^\perp_{\Gamma_1}}$ spans the whole vector space. Using Eqs.~\eqref{complat}, \eqref{gformula}, and \eqref{tcom}, we have
\eqn{
g_{\Gamma_{1}}=\sqrt{v_{0}\left({\cal T}\right)}=\frac{1}{4K}.
}
Similar to the above case, the fully localized field has two possible solutions corresponding to $\theta_{1}=\theta_{2}=\tilde{\theta}_{1}=\tilde{\theta}_{2}=0$ or $\theta_{1}=\theta_{2}=\tilde{\theta}_{1}=\tilde{\theta}_{2}=\pi$.  We thus have the additional degeneracy $d=2$. Taken together, we obtain the $g$ function of the $K>1/2$ TLL under phase measurement by
\eqn{\label{gbitflip}
g_{{\cal E}}=\underbrace{(4K)^{-1}}_{g_{\Gamma_{1}}}\times\underbrace{1}_{g_{\Gamma_{2}}}\times\underbrace{2}_{d}=\frac{1}{2K},
}
which provides Eq.~\eqref{gfuncbit}.

\section{Numerical results\label{sec:num}}
Below we numerically test the field-theoretical results obtained above. Specifically, we consider the spin-$\frac{1}{2}$ XXZ chain described by the following Hamiltonian:
\eqn{\label{xxzh}
\hat{H}_{{\rm XXZ}}=J\sum_{i=1}^L\left(\hat{\sigma}_{i}^{x}\hat{\sigma}_{i+1}^{x}+\hat{\sigma}_{i}^{y}\hat{\sigma}_{i+1}^{y}+\Delta\hat{\sigma}_{i}^{z}\hat{\sigma}_{i+1}^{z}\right),
}
where $J>0$ and the periodic boundary condition is imposed on the spatial direction. When $|\Delta|<1$, it is well-known that its ground state is gapless and described by the TLL with $K\ge1/2$. In particular, the relation between the TLL parameter $K$ and the anisotropy $\Delta$ has been obtained from the Bethe ansatz as follows:
\eqn{\label{bethe}
K=\frac{\pi}{2\left(\pi-\cos^{-1}\Delta\right)}.
}
We perform the exact diagonalization of $\hat{H}_{{\rm XXZ}}$ by using Lanczos algorithm and calculate the ground-state wavefunction $|\Psi_0\rangle$. 
To obtain the vector representation $|\rho_{\cal E}\rparen$ of the reduced system density matrix in the doubled Hilbert space as in Eq.~\eqref{rhoed2}, we first construct the tensor product $|\Psi_0\rangle\otimes|\Psi_0\rangle$. An imaginary-time evolution $e^{-\mu[\sum_j(1-\hat{\sigma}_j^\alpha\otimes\hat{\sigma}_j^\alpha)]}$ is then acted on it, where $\alpha$ is chosen to be either $z$ or $x$ depending on whether we consider density or phase measurement.  We numerically evaluate the R\'enyi entropy $S_{SE}$ in Eq.~\eqref{ssez} by calculating the norm of $|\rho_{\cal E}\rparen$. Finally, we determine the universal contribution $s_0$ by fitting $S_{SE}$ to a scaling form
\eqn{
S_{SE}=s_{1}L-s_{0}+\frac{s_{-1}}{L}.
}
Despite relatively small system sizes available in the exact diagonalization, our numerical analysis confirms the analytical predictions with high accuracy as we discuss now.

\begin{figure}[b]
\includegraphics[width=80mm]{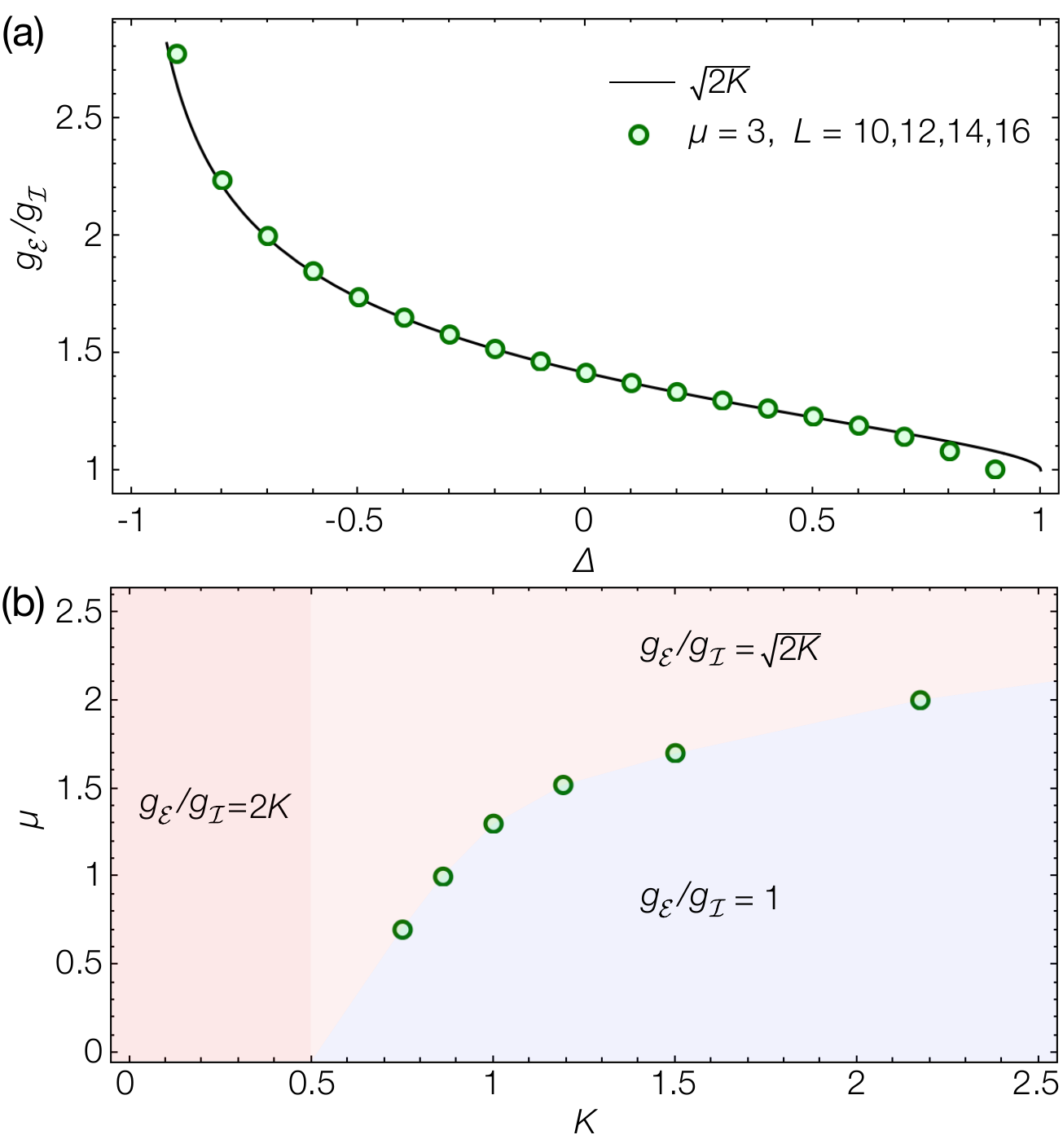} 
\caption{\label{fig_z}
Universal contribution $e^{s_0}=g_{\cal E}/g_{\cal I}$ and the phase diagram of the TLL under density measurement.
(a) Numerical values of the $g$ functions at different $\Delta$. The measurement strength is chosen to be $\mu=3$, at which the values of the $g$ functions are converged. The system sizes used for the fitting are $L=10,12,14,16$. The solid curve shows the analytical prediction $\sqrt{2K}$ with $K$ given by Eq.~\eqref{bethe}. (b) Phase diagram in the space of the TLL parameter $K$ and the measurement strength $\mu$. Numerical values of the critical strengths $\mu_c$ are extracted from the data collapses in Fig.~\ref{fig_z_g}(d)-(f). We note that only the case of $K\ge 1/2$ can be investigated in the spin-$\frac{1}{2}$ XXZ chain~\eqref{xxzh}. 
}
\end{figure}

\subsection{Density measurement\label{subsec:numz}}
We first present the results in the case of the $K>1/2$ TLL under density measurement, for which the nonperturbative RG analysis predicts the transition occurring at a nonzero measurement strength $\mu_c$. As discussed in Sec.~\ref{subsec:rgz} this boundary phase transition originates from the anomalous enhancement of the $k^2$ kinetic term in the boundary action ${\cal S}_{\cal E}$, which effectively reduces the value of $K$ near the boundary and tends to render the boundary state susceptible to the perturbations. The resulting RG flows exhibit the nonmonotonic behavior leading to the  locking of $\varphi_-$ in the IR limit (cf. Fig.~\ref{fig_rg}(a)). 
 From the boundary CFT analysis, we find that the universal contribution $e^{s_0}=g_{\cal E}/g_{\cal I}$ should discontinuously change from $1$ to $\sqrt{2K}$ across the transition. 

\begin{figure}[b]
\includegraphics[width=80mm]{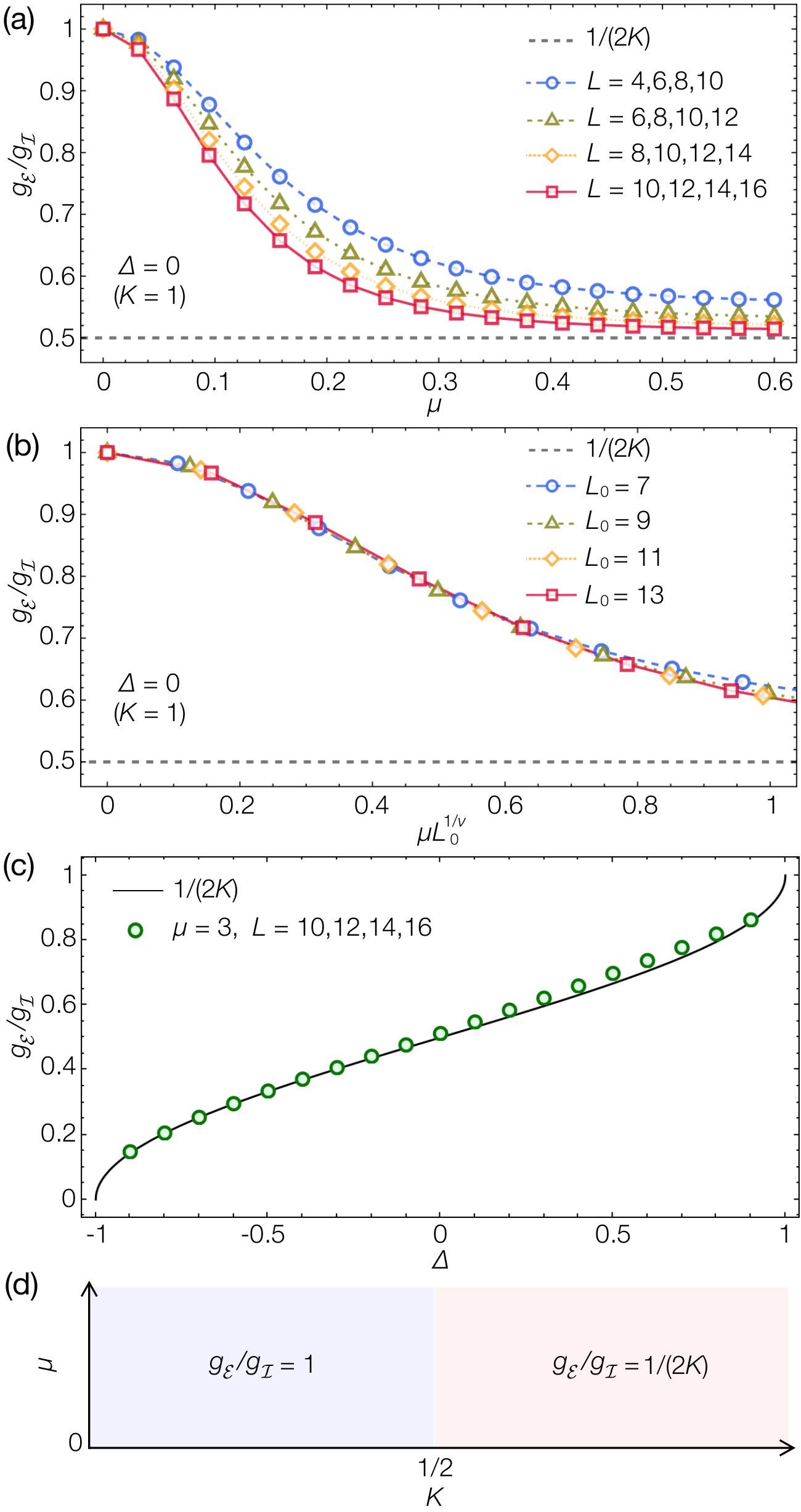} 
\caption{\label{fig_x}
Universal contribution $e^{s_0}=g_{\cal E}/g_{\cal I}$ and phase diagram of the TLL under phase measurement. 
(a) Numerical values of the $g$ function extracted from the second-order R\'enyi entropy fitted to $s_{1}L-s_{0}+\frac{s_{-1}}{L}$ with $L\in\{L_{0}-3,L_{0}-1,L_{0}+1,L_{0}+3\}$ and $L_0=7,9,11,13$ at $\Delta=0$, and (b) the corresponding data collapse which yields $\nu\approx 1.6$. (c) Converged values of the $g$ function at $\mu=3$ compared with the analytical prediction $1/(2K)$ indicated by the solid curve.  (d) Phase diagram in the space of the TLL parameter $K$ and the measurement strength $\mu$. 
}
\end{figure}

These results are numerically checked in Fig.~\ref{fig_z_g}, where the estimated universal contributions at different system sizes and various $\Delta$ are plotted against the measurement strength $\mu$ in (a)-(c). Interestingly, the $g$ function grows as the coupling $\mu$ is increased and then eventually converges to a value of $\sqrt{2K}$ as predicted by the boundary CFT analysis in Eq.~\eqref{gs2k}.  The corresponding data collapses are shown in Fig.~\ref{fig_z_g}(d)-(f) which are obtained by assuming the scaling form
\eqn{
\frac{g_{{\cal E}}}{g_{{\cal I}}}=f\left(\left(\mu-\mu_{c}\right)L_{0}^{1/\nu}\right),
}
where we find that the exponent $\nu\approx 6.0$ fits well the numerical data. 

Figure~\ref{fig_z}(a) shows the converged values of the universal contribution as a function of the anisotropy $\Delta$. These results show remarkable agreement over a broad range with the analytical prediction $\sqrt{2K}$. Figure~\ref{fig_z}(b) plots the phase diagram in the space of the TLL parameter $K$ and the measurement strength $\mu$, where the transition points in $K>1/2$ are extracted from the data collapses as done in Fig.~\ref{fig_z_g}(d)-(f). We find that a threshold value $\mu_c$ monotonically increases as a function of $K$, which is qualitatively consistent with our RG analysis.  These results suggest that, in contrast to what is expected from a perturbative analysis, the system-environment entanglement  can exhibit the universal phase transitions as a function of the measurement strength $\mu$.

\subsection{Phase measurement\label{subsec:numx}}
We next present the numerical results for the $K>1/2$ TLL under phase measurement.  
Figure~\ref{fig_x}(a) shows the universal contribution $e^{s_0}=g_{\cal E}/g_{\cal I}$ plotted against the measurement strength $\mu$ at $\Delta=0$ and different system sizes. 
The $g$ function monotonically decreases as a function of $\mu$ this time and eventually converges to a value close to the analytical prediction $1/(2K)$ indicated by the dashed horizontal line. 
The corresponding data collapse is shown in Fig.~\ref{fig_x}(b), where we assume the scaling form
\eqn{
\frac{g_{{\cal E}}}{g_{{\cal I}}}=f\left(\mu L_{0}^{1/\nu}\right).
} 
The positivity of the exponent $\nu$ indicates that the phase measurement acts as a relevant perturbation to the TLL at $\mu>0$. The monotonic decrease of the universal contribution implies that the $g$ function behaves as the RG monotone as expected from the $g$-theorem.  

Figure~\ref{fig_x}(c) compares the converged values of the $g$ function at different $\Delta$ with the analytical prediction $1/(2K)$ in Eq.~\eqref{gbitflip}.  
We again find the agreement between the two results with an error below $\sim 5\%$ over a broad range of the parameter.
The phase diagram of the TLL under phase measurement is shown in Fig.~\ref{fig_x}(d), in which the phase boundary is described by the vertical line independent of the measurement strength $\mu$ \cite{Kane92,WU12}. We note that the difference between Fig.~\ref{fig_z}(b) and Fig.~\ref{fig_x}(d) is due to the lack of the dangerously irrelevant term in the latter.

\section{Possible experimental realization\label{sec:exp}}
We briefly discuss a possible way to experimentally test our theoretical predictions. To be concrete, we propose an ultracold atomic experiment on the lines of previous studies \cite{DAJ12,IR15}. Our main interest lies in measuring the second-order R\'enyi entropy for the post-measurement density matrix of the entire system (see Eq.~\eqref{secondrenyi}). We emphasize that, in contrast to the entanglement measures in studies of measurement-induced phase transitions,  the quantity of interest to us requires no postselections since it is defined for the unconditioned nonunitary evolution. The following is our proposal for measuring the system-environment entanglement in the TLL under density measurement:
\begin{itemize}
\item[(i)]{Prepare the two identical copies of a 1D critical Bose gas described by the TLL. This can be naturally realized in ultracold atomic experiments \cite{HS08,IR15,YB17}. After the preparation, the unitary dynamics is frozen by, e.g., rapidly increasing the depth of an optical lattice and switching off interactions by Feshbach resonances.}
\item[(ii)]{Perform a weak density measurement while discarding the outcomes. This induces a controlled decoherence on the two copies. In practice, such process can be realized by shining a probe light on ultracold gases which leads to light scattering in a similar manner as, for instance, routinely done in quantum gas microscopes \cite{BWS09,SJF10}. Technically, such process can be described by the Markovian master equation~\eqref{liouville} whose jump operator is given by a site-resolved occupation number $\hat{L}_j=\sqrt{\gamma}\hat{n}_j$ \cite{PH10,YA15}. The measurement strength $\mu=\gamma t$ can be controlled by changing either the exposure time $t$ or the intensity of the probe light that determines the scattering rate $\gamma$ \cite{LHP17,SF20}.}
\item[(iii)]{Perform a beam-splitter operation between the two copies. This can be achieved by lowering the potential barrier between the chains and letting the atoms tunnel between the two copies for a certain duration of time.}
\item[(iv)]{Perform the projection measurement on the two copies to determine the site-resolved occupation number $\{n_{j,\alpha}\}$ in each copy $\alpha\in\{1,2\}$. In practice, this can be realized again by quantum gas microscopes. The second-order R\'enyi entropy of the entire system in Eq.~\eqref{secondrenyi} can then be obtained by evaluating the expectation value of the swap operator $(-1)^{\sum_j n_{j,2}}$ after repeating the whole procedures.}
\end{itemize}

While measurements performed in steps (i) and (iii) are technically the same type of processes corresponding to light scattering, the key point in our proposal is that the measurement strengths of them can be quite different. In step (iii), one typically requires the use of near-resonant probe light to realize high scattering rate and a clear fluorescence image, allowing one to determine the occupation number with almost unit fidelity \cite{BWS09,SJF10}. In step (i), however, there is no such need since the measurement outcomes will be discarded anyway; to realize a less destructive controlled decoherence, one can use an off-resonant or low-intensity probe light whose wavelength does not even need to be comparable to the lattice constant \cite{YA15,PYS14,PYS15}. Also, the periodic boundary conditions might be realized by a ring-shape optical potential \cite{AL22} or by utilizing a programmable platform such as a Rydberg atom array \cite{bluvstein2022quantum}. 
Our numerical results suggest that a relatively small system with tens of lattice sites should be enough to test the universal behavior of the system-environment entanglement. As such, we expect that our theoretical predictions are within reach of current experimental techniques. 

We note that, under the open boundary conditions, the subleading contribution to $S_{SE}$ is given by the logarithmic term $-s'\ln(L)$ whose coefficient $s'$ can exhibit a universal behavior depending on the TLL parameter $K$ \cite{SJM11,ZMP11}. We expect that the entanglement phase transition, which occurs as a function of the coupling strength $\mu$ (cf. Fig.~\ref{fig_z}(b)), can be observed also in such open-boundary systems as a singular change of the universal coefficient $s'$; we will address this issue in detail in a future work.

From a broader perspective, we also mention that a circuit QED setup might give another route toward testing some of our predictions \cite{BA21}.  The TLL has been experimentally realized in, for instance, the long superconducting waveguide consisting of Josephson junctions, where the TLL parameter can be controlled by changing the wave impedance \cite{KR19}. When coupled to an impurity Josephson junction, the low-energy physics can be described by the same type of the effective action analyzed in Sec.~\ref{sec:rg}. Notably, the corresponding boundary behavior has been experimentally studied by measuring phase shifts in microwave photon scatterings \cite{KR23}. Further developments of these techniques might allow one to directly confirm a variety of the conformal boundary conditions discussed in this paper.

\section{Summary and Discussions\label{sec:sum}}
We have studied the universal aspects of the entanglement inherent to open many-body systems, i.e., the entanglement between a system of interest and its environment. We have demonstrated that a TLL under a local measurement can exhibit a universal entanglement phase transition when the measurement strength is varied. We emphasize that this occurs  in the unconditioned evolution, where the outcomes are averaged over and no postselections are necessary. The universality of the system-environment entanglement is encoded in the size-independent contribution to the R\'enyi entropy of the post-measurement density matrix. We have determined the value of the universal term by developing the field-theoretical formalism in the doubled Hilbert space and applying the boundary CFT techniques to the multicomponent field theory. The results have been verified by the numerical calculations in the spin-$\frac{1}{2}$ XXZ chain. Finally, we have discussed a possible way to test our theoretical predictions in ultracold atomic experiments.

Several interesting directions remain for future studies. First, our field-theoretical formalism is not specific to the problem considered in this paper  but can be applied to a variety of settings. One natural direction, for instance, is to analyze the entanglement between subregions within a system subject to the influence of the environment. This might provide realization of a many-body analogue of the so-called environment-induced sudden death of entanglement \cite{TY09}, in which a highly entangled many-body state would become a product state at a nonzero but finite strength of the system-environment coupling. One can also consider a situation in which only a subregion of the system is influenced by the environment \cite{MAR16}. 

Second, it merits further study to extend the present analysis to the case in which the Kraus operators cannot be diagonal in a field basis. We recall that in the present work the Kraus operators can be treated as diagonal variables in either $\phi$ or $\theta$ representation, leading to the boundary action localized at $\tau=0$ in the Euclidean path integral. In contrast, the effect of nondiagonal Kraus operators should be expressed as an action that is not strictly localized in the imaginary-time direction $\tau$. Similarly, while the unitary dynamics is assumed to be frozen during the measurement process in the present work, its inclusion might further enrich entanglement structures. It remains open how one could generalize boundary CFT techniques to those situations if at all possible. It would also merit further study to identify which type of measurement leads to conformally invariant boundary conditions; it has been found that some measurements can impose non-conformally invariant boundary conditions \cite{NK16,HM24,ZY23}.

Finally, it is intriguing to ask how one could experimentally test our theoretical predictions. In the present work, we have proposed a concrete protocol for this, which we believe to be within reach of current ultracold atomic experiments.  It merits further study to explore the possibility of studying the system-environment entanglement in another quantum platform, especially in view of recent developments of programmable quantum devices. The spin-$\frac{1}{2}$ XXZ chain, for instance, has been realized in superconducting quantum processors \cite{MA22}, and its entanglement structure should be diagnosed by performing local random measurements followed by a certain classical postprocessing \cite{EA20}.  One possible challenge here is an implementation of a well-controlled decoherence on programmable quantum platforms.   
We hope that our work stimulates further studies in these directions.

\begin{acknowledgments}
We are grateful to Yohei Fuji, Hosho Katsura, Kanta Masuki, Keiji Saito, Takeru Yokota, and Tsuneya Yoshida for useful discussions. Y.A. acknowledges support from the Japan Society for the Promotion of Science (JSPS) through Grant No. JP19K23424 and from JST FOREST Program (Grant Number JPMJFR222U, Japan) and JST CREST (Grant Number JPMJCR23I2, Japan). S.F. acknowledges support from JSPS through Grant No. JP18K03446 and from the Center of Innovations for Sustainable Quantum AI (JST Grant No. JPMJPF2221).  The work of M.O. was partially supported by the JSPS KAKENHI Grants No. JP23K25791 and No. JP24H00946.
\end{acknowledgments}

\appendix
\section{Details about the nonperturbative RG analysis\label{app:frg}}
We provide technical details about the fRG analysis performed in Sec.~\ref{subsec:rgz}. 
We consider the theory described by the action $\cal S$ in Eq.~\eqref{Sdephasing} and aim to determine its low-energy behavior. In fRG, we use the effective action $\Gamma_\Lambda$ at energy scale $\Lambda$  that interpolates the bare action at a UV scale $\Gamma_{\Lambda_0}={\cal S}$ and the effective action $\Gamma_0=\Gamma$ in the IR limit with $\Gamma$ being the generating functional of one-particle irreducible correlation functions. 
The flow equation of  $\Gamma_\Lambda$ is given by the exact RG equation, which is hard to solve without making any approximations \cite{DN21}.  

To make the calculations tractable, we make several simplifications. First of all, as discussed in the main text, we neglect the cross coupling term $\cos(\varphi_+)\cos(\varphi)_-$ which is less relevant compared to the other potential perturbations. The action can then be decoupled into the two sectors including only either of $\varphi_+$ or $\varphi_-$. The ground-state properties in the $+$ sector have been well-understood from the perturbative RG analysis and the duality argument \cite{Kane92}. We here focus on the fRG analysis of the $-$ sector.  Specifically, we assume the following LPA' ansatz:
\eqn{\label{fansatz}
\Gamma_{\Lambda}[\varphi]&=&\frac{1}{2}\int_{-\infty}^{\infty}\frac{dk}{2\pi}\left(\frac{|k|}{4\pi K}+\frac{\gamma k^{2}}{\Lambda}\right)|\varphi_{k}|^{2}\nonumber\\
&&-u\Lambda\int_{-\infty}^{\infty}dx\,\cos\left(\varphi(x)\right),
} 
where we abbreviate the subscript $-$ in $\varphi$ and $u$ for the sake of notational simplicity. We then consider the following exact RG equation within this functional ansatz:
\eqn{\label{erg}
\Lambda\partial_{\Lambda}\Gamma_{\Lambda}=\frac{1}{2}{\rm Tr}\left[\partial_{\Lambda}R_{\Lambda}G_{\Lambda}\right],
}where we choose the regulator to be $R_{\Lambda}=\Lambda\frac{k/\Lambda}{e^{k/\Lambda}-1}$, which allows for relatively simple expressions of the beta functions as shown below.  The propagator $G_\Lambda$ is defined as
\eqn{
G_{\Lambda}(y)\!=\!\frac{1}{\frac{|y|}{4\pi K}+\gamma y^{2}+u\cos(\varphi)+\frac{y}{e^{y}-1}},\;\;y=\frac{k}{\Lambda}.
}
The flow equations of the parameters $u$ and $\gamma$ can be obtained by projecting Eq.~\eqref{erg} onto the ansatz~\eqref{fansatz}. The results are
\eqn{
(1+\Lambda d_{\Lambda})u\!&=&\!-\frac{1}{2\pi^{2}}\!\int_{0}^{2\pi}d\varphi\cos\left(\varphi\right)\!\!\int_{0}^{\infty}\!\!dy\frac{y^{2}G_{\Lambda}(y)}{4\sinh^{2}\left(y/2\right)},\nonumber\\ \\
\left(-1+\Lambda d_{\Lambda}\right)\gamma&=&\frac{u^{2}}{4\pi^{2}}\int_{0}^{2\pi}\sin^{2}(\varphi)\int_{0}^{\infty}dy\frac{y^{2}G_{\Lambda}^{2}(y)}{4\sinh^{2}\left(y/2\right)}\nonumber\\
&&\times\lim_{y'\to0}\partial_{y'}^{2}G_{\Lambda}\left(y+y'\right).
}
Finally, after performing the integrations over the phase variable $\varphi$, we can derive the beta functions for each parameter by
\begin{widetext}
\eqn{
\beta_{u}(u,\gamma)&=&u-\int_{0}^{\infty}dy\frac{y^{2}}{4\pi u\sinh^{2}\left(y/2\right)}\left(\frac{\zeta}{\sqrt{\zeta^{2}-u^{2}}}-1\right),\;\;\;\zeta(y)\equiv\frac{y}{4\pi K}+\gamma y+\frac{y}{e^{y}-1},\\
\beta_{\gamma}(u,\gamma)&=&-\gamma+u^{2}\int_{0}^{\infty}dy\frac{y^{2}\left[\left(\left(\partial_{y}^{2}\zeta\right)\zeta-2\left(\partial_{y}\zeta\right)^{2}\right)\left(u^{2}+4\zeta^{2}\right)-5u^{2}\zeta\partial_{y}^{2}\zeta\right]}{64\pi\sinh^{2}\left(y/2\right)\left[\zeta^{2}-u^{2}\right]^{7/2}},
}
\end{widetext}
which give the full expressions of Eq.~\eqref{betafunc} in the main text.

We note that the correlation function $C(x)$ plotted in Fig.~\ref{fig_beta}(a) can be obtained by the Fourier transform of the quadratic part of the propagator, $G^0_{\Lambda}(y)=[|y|/(4\pi K)+\gamma y^2]^{-1}$, leading to the following expression:
\begin{widetext}
\eqn{
   \log C(x) \!=\! 
    2 K \left(2 \text{Ci}\left(\frac{| x| }{2a \gamma  K }\right) \!\cos \left(\frac{| x|
    }{2a   \gamma  K}\right)\!+\!2 \text{Si}\left(\frac{| x| }{2a \gamma  K  }\right)\! \sin
    \left(\frac{| x| }{2a   \gamma  K}\right)\!-\!\pi  \sin \left(\frac{| x| }{2a  \gamma
     K}\right)\!-\!2 \log (| x|)\! \right)\!+\!\text{const.}, \label{corrc}
}
\end{widetext}
where $\mathrm{Ci}$ and $\mathrm{Si}$ are cosine and sine integral, respectively. In the absence of the boundary term $\gamma$, Eq.~\eqref{corrc} reproduces the well-known critical decay $C(x)\propto1/|x|^{4K}$.

\newpage

\bibliography{reference}

\end{document}